\documentclass[aps,jcp,twocolumn,groupedaddress
 reprint,
superscriptaddress,
showpacs,preprintnumbers,
 amsmath,amssymb,
 aps,
]{revtex4} 
\usepackage{graphicx}
\usepackage{dcolumn}
\usepackage{bm}%

\begin{document}
\newcommand {\be}{\begin{equation}}
\newcommand {\ee}{\end{equation}}
\newcommand {\bea}{\begin{array}}
\newcommand {\cl}{\centerline}
\newcommand {\eea}{\end{array}}
\newcommand {\pa}{\partial}
\newcommand {\al}{\alpha}
\newcommand {\de}{\delta}
\newcommand {\ta}{\tau}
\newcommand {\ga}{\gamma}
\newcommand {\ep}{\epsilon}
\newcommand {\si}{\sigma}
\newcommand{\up}{\uparrow}
\newcommand{\down}{\downarrow}

\title{Charge transport in doped zigzag phosphorene nanoribbons}

\author{Zahra Nourbakhsh}
\affiliation{School of Nano Science, Institute for Research in Fundamental Sciences (IPM), Tehran 19395-5531, Iran}
\author{Reza Asgari}
\affiliation{School of Nano Science, Institute for Research in Fundamental Sciences (IPM), Tehran 19395-5531, Iran}
\affiliation{School of Physics, Institute for Research in Fundamental Sciences (IPM), Tehran 19395-5531, Iran}

\begin{abstract}

The effects of lattice distortion and chemical disorder on charge transport properties of two-terminal zigzag phosphorene nanoribbons (zPNRs), which shows resonant tunneling behavior under an electrical applied bias, are studied. Our comprehensive study is based on {\it ab~initio} quantum transport calculations on the basis of the Landauer theory. We use nitrogen and silicon substitutional dopant atoms, and employ different physical quantities such as the $I-V$ curve, voltage drop behavior, transmission spectrum, transmission pathway, and atomic current to explore the transport mechanism of zPNR devices under a bias voltage. The calculated transmission pathways show the transition from a ballistic transport regime to a diffusive and in some particular cases to localized transport regimes. Current flowing via the chemical bonds and hopping are monitored, however, the conductance originates mainly from the charge traveling through the chemical bonds in the vicinity of the zigzag edges. Our results show that, in the doped systems, the device conductance decreases and the negative differential resistance characteristic becomes weak or is eliminated. Besides, the conductance in a pure zPNR system is almost independent of the ribbon width.
\end{abstract}

\maketitle

\section{Introduction}\label{sec:intro}

Nowadays, low-dimensional electronic devices have gained great attention owing to their particular properties which could lead to a revolution in technology  \cite{2Dpot}. Graphene is the most well-known nanostructure material; however, the absence of a band gap has limited its application in electronic devices.  During the last couple of years, researchers have looked for other two-dimensional (2D) crystalline materials beyond graphene nanomaterials, and monolayer phosphorene has been discovered as a promising system.  It is an elemental 2D crystalline material where its bulk structure, named black phosphorus, is highly stable in ambient conditions. Similarly to graphite, 2D phosphorene layers in black phosphorus are kept together by van der Waals forces. The main properties of desirable electronic devices have been gathered in phosphorene. It has a $\sim1.4$~eV direct band gap, high carrier mobility and high on/off ratio for field-effect transistor applications \cite{ph-birth, ph-appl1, ph-appl2, ph-appl3, aniso1}. In addition, the remarkable characteristic of phosphorene as a 2D crystalline system, is its high in-plane anisotropic properties which arises from its structure and lead to the appearance of anisotropic behavior in its electronic structure, mechanical properties, optical absorption, electrical and thermal conductivity, transport, and optoelectronic responses \cite{aniso1, aniso2, PExcExp,asgari}.

As a general step toward the further investigation of phosphorene-based nanoelectronic devices, phosphorene nanoribbons (PNRs) have been extensively studied \cite{nour,pnrs,pnrtran1,pnrtran2,pnrtran3,pnrtran4}. According to the PNR edge structures, they are elementarily classified into two types, namely armchair PNRs (aPNRs) and zigzag PNRs (zPNRs). It has recently been shown by us that the charge distributions around the gap in aPNRs are localized in the middle of the nanoribbons, while in zPNRs they are mostly distributed in the width of the ribbons \cite{nour}. Consequently, physical properties of zPNRs are very sensitive to the width, structural defects, and edge morphologies. Therefore, they provide additional ways to access new physics in zPNR devices \cite{nour,pnrtran3}, and open up a door for further research to explore new design with an experimentally controllable manner.

It is worth mentioning that the edge-terminated zPNRs and aPNRs are semiconductors with a direct band gap at the $\Gamma$ point \cite{nour}. However, bare aPNRs (aPNRs with free edges) are semiconductors with an indirect band gap, and in addition, bare zPNRs (zPNRs with free edges) are metals \cite{pnrtran2}.  There are some theoretical studies on transport properties of the bare and edge-terminated PNR systems \cite{pnrtran1, pnrtran2, pnrtran3,pnrtran4} in which a high current intensity and non linear $I-V$ behavior, named negative differential resistance (NDR), where current decreases through the bias voltage enhancement, are predicted. zPNRs are the only known homostructure devices which show NDR \cite{pnrtran1}. This particular property is important in high-speed switches, logic devices, amplifiers, multi-valued memories, and high-frequency oscillators \cite{ndr}. Besides, low-bias NDR devices are useful to reduce power consumption of electronic devices \cite{ndr}.

Knowing the sensitivity and possible tunability of zPNR physical properties to structural manipulations and imperfections, we can investigate the influence of structural distortion and chemical defects on transport properties of the zPNR devices. Such a study is very essential to design experimental structures. For this purpose, we consider nitrogen (N) and silicon (Si) substitutional dopants in a zPNR system. Both N and Si atoms are phosphorus neighbors in the periodic table, and respectively as kinds of isovalence and isosize impurities are good candidates to tune the transport properties of zPNRs; N and Si are also two of the most abundant elements in the environment.
Substitution of Si by phosphorus atoms in silicon devices is a well-known subject in semiconductor technology as a way to enhance transport properties of the system. Here, we are interested in exploring the effect of Si substitutional dopant in transport properties of the zPNRs. Si dopant atoms could induce different changes on the electronic structure of systems around the Fermi level which will be discuss in Sec. \ref{sec:struct}.
Besides, we should note that owing to the puckered structure of phosphorene, the appearance of different kinds of lattice defects and structural distortions in phosphorene compared to other 2D materials is more probable (for instance see Ref. \cite{distP}). Considering the N substitutional dopant by conserving the $sp^3$ hybridization of the lattice gives us the opportunity to explore the impact of the lattice distortion on transport properties of the system.

In this article, we consider different decoration of dopant atoms in the free-edges zPNR background to reveal the transport sensitivity to disorder place.
Since some important features of charge transport in zPNRs are not clear, we will study how charge carriers travel through the device. Moreover, the voltage drop behavior, the local current density in pure zPNR devices, and the effect of the ribbon width on transport behavior are discussed.

We carry out {\it ab~initio} electronic calculations, using the combination of density functional theory (DFT) \cite{k-sh} with nonequilibrium Green's function (NEGF) \cite{negf} methods based on the successful Landauer formalism \cite{landauer}, which explains the carrier transport through mesoscopic systems in terms of scattering process.
This method has been widely used to calculate the transport properties of different classes of mesoscopic systems and good agreements between experiment and theory have been reported \cite{exp1,exp2,exp3,exp4,exp5}.

The rest of the paper is organized as follows. The theoretical and technical methods are discussed in Sec. \ref{sec:method}. We present our results in Sec. \ref{sec:results}; namely, a brief description of the electronic structure of pure and doped 2D phosphorene, and zPNRs, calculating transport properties in equilibrium and nonequilibrium conditions for different devices, the $I-V$ characteristic, the transmission spectrum, the carrier types and their transmission pathway, local current density, and voltage drop behavior of the proposed systems are presented. Finally, we end with a brief summary of our main results in Sec. \ref{sec:summary}.

\section{Methodology}\label{sec:method}

\begin{figure}
\centering
\includegraphics[width=1\linewidth]{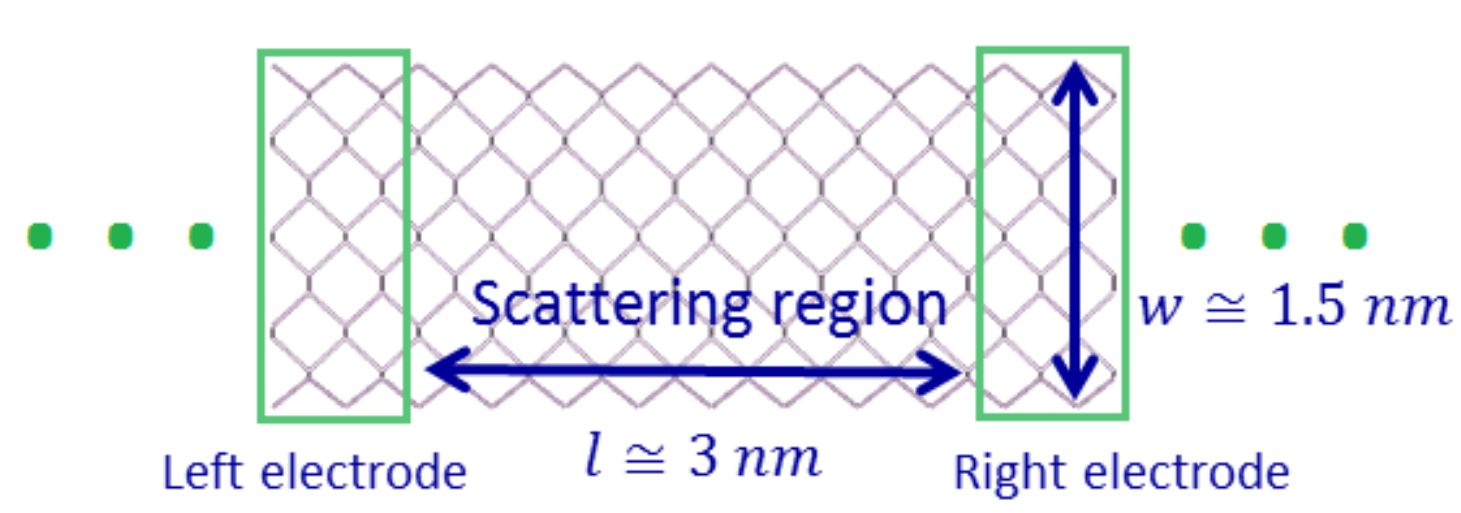}
\caption{\label{device} (Color online) The computational setup for a two-terminal free-edge zPNR device. The system consists of three parts: semi-infinite left and right electrodes, and the central scattering region. The length and width of the scattering region are shown in the picture. The dots denote the semi-infinite directions. The boxes show the left and right electrode supercells containing two primitive unit cells with the length of $\sim 0.7$~nm. The electrode size in the transport direction should be large enough that its orbitals only interact with the nearest neighbor cell in the repeating system.}
\end{figure}

A schematic representation of a two-terminal free-edges zPNR device is shown in Fig.~\ref{device}. As illustrated in the figure, this system consists of three main parts involving semi-infinite left and right electrodes, and the finite central scattering region. According to the semi-infinite character of electrodes, the electrode states are not perturbed by passing and reflecting the charge carriers. The scattering region is considered to be large enough to avoid any overlap between the wave functions of the right and the left electrodes. Under the bias voltage, the electrodes are at two different electrochemical potentials. This causes that the charge carriers from the source electrode, are partially transmitted through the central region, and finally absorbed in the drain electrode; also the reflected carriers come back to the source electrode. Accordingly, the system is fundamentally different from the standard periodic systems. We face an open boundary system along the transport direction with a nonequilibrium electron density in the central region. To describe this system, we thus use Green's function techniques on the basis of the Landauer formula, which explains the quantum transport process using the scattering theory \cite{landauer}. Also, this gives rise to assuming only coherent transport of carriers through the device.

To compute the transport properties of system, we use the combination of $ab~initio$ density functional methods with non-equilibrium Green's function techniques (DFT-NEGF) implemented in the TranSIESTA software \cite{transiesta}. This code provides the ability of calculating the electron density of an open boundary system under a possible applied bias by self-consistent solution of the Poisson equation with appropriate boundary conditions. In addition to the charge density, the electrostatic Hartree potential and Kohn-Sham effective Hamiltonian are obtained by the TranSIESTA calculations.

The first step to evaluate the transport properties is to calculate the transmission spectrum, which depends on the energy of the incoming carriers from the source electrode and bias voltage, $T(E,V_{Bias})$. In order to understand and analyze the transport mechanism, the transmission spectrum can be integrated, decomposed or projected in different ways to extract the related quantities such as current intensity, transmission pathway, and current density \cite{handbook}.  According to the standard Landauer formalism, current is obtained by taking the integral of the transmission coefficient over the bias energy window. The relation at bias voltage $V$ is given by
\be
\label{landaue}
I(V)=\frac{G_0}{e}\int_{\mu_L}^{\mu_R}T(E,V)dE
\ee
where $e$ and $\mu_{L(R)}=E_{\rm F}\pm eV/2$ are the electron charge and the electrochemical potentials of the left (right) electrode, respectively. The difference of the electrode potentials is basically given by the value of the bias voltage, $eV=\mu_L-\mu_R$. The coefficient $G_0=2e^2/h$ ($h$ is the Planck constant) is known as the quantum of the conductance. The transmission coefficient is obtained from the retarded Green's function using $T(E,V) =  \Gamma_L(E,V) G(E,V) \Gamma_R(E,V) G^{\dag}(E,V)$, where $\Gamma_{L(R)}$ is the level broadening defined by $\Gamma_{L(R)}=i(\Sigma_{L(R)}-\Sigma^{\dag}_{L(R)})$, $\Sigma_{L(R)}$ is the self-energy which describes the coupling of the semi-infinite left (right) electrode to the finite central region, and the retarded Green's function of the central region is computed by
$G=[E+i\eta - H - \Sigma_L - \Sigma_R]^{-1}$, where $H$ is a Hamiltonian matrix of the central region.

Before computing the transport properties, on the basis of $ab~initio$ methods, the geometries of the studied structures are relaxed in
order that all components of all forces become less than $0.01$ eV/\AA. The geometry optimization and electronic structure calculations
are performed using the SIESTA code \cite{siesta} based on $ab~initio$ DFT methods, and the transport calculations are carried out using TranSIESTA. We use nonlocal Troullier-Martins norm-conserving pseudopotentials to describe the core electrons and linear combinations of the atomic orbitals to expand the valence electrons. A double-$\zeta$ polarized basis set within the generalized gradient approximation (GGA) in the scheme of Perdew, Burke, and Ernzerhof (PBE) \cite{pbe} is used in our calculations.
The real space Fourier expansion of the electron density is cut at $150$~Ry. The Brillouin zone integrations are performed on the
Monkhorst-Pack \cite{m-p} $k$-point grid of $1 \times 1 \times 10$, for the electronic structure calculations. It increases to
$1 \times 1 \times 70$ during the transport calculations. Note that such a large number of $k$ points is needed to simulate the semi-infinite electrodes.
To avoid the unphysical interactions between periodic images and simulate an isolated system, we use a vacuum gap of $20$~\AA.
The electron temperature is set at $300$~K in our transport calculations. Our studied structures do not show any considerable magnetic moment. 
As our calculations are limited to the low biases, we ignore the computationally difficult structural relaxations induced by the bias voltage.

In addition, we use several post-processing codes such as TBtrans (tight-binding transport)\cite{tbtran} and the sisl utility \cite{sisl}, as well as some home made programming to drive and extract the electronic transport properties of the system.

\section{RESULTS and DISCUSSION}\label{sec:results}

Electronic band structure could provide some useful understanding of transport behaviors around the zero bias voltage. Thus, before studying the transport properties, we investigate the effect of Si and N substitutional doping on the electronic structure of 2D phosphorene and zPNR systems. Notice that the values of the length and width of the scattering device are $\sim 3$ and $1.5$ nm in all figures, unless we specifically define those values otherwise.

\subsection{Geometric and electronic structures of doped zPNRs}\label{sec:struct}

\begin{figure}
\centering
\includegraphics[width=0.9\linewidth] {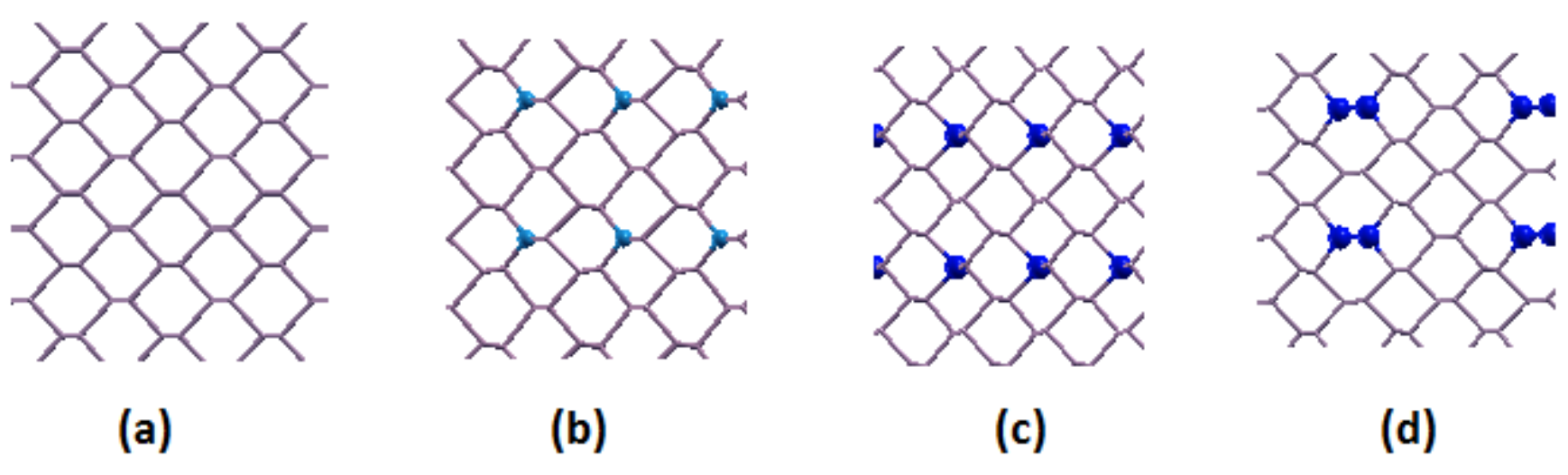}
\includegraphics[width=1\linewidth] {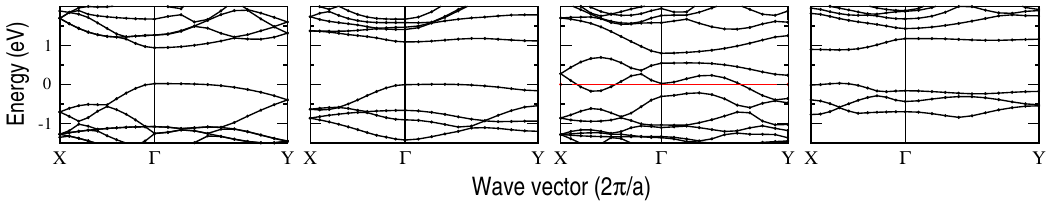}
\caption{\label{2dStruct} (Color online) Top row: The top view of 2D phosphorene (a), the N-doped phosphorene system (b), and two types of Si-doped systems (c) and (d). The impurity concentrations in the doped systems are the same. N atoms are in cyan, and Si atoms (the larger balls) are in blue.
Bottom row: The electronic band structure diagrams of the related systems in similar-size supercells. In the semiconductor (metallic) systems, the top of the valence bands (Fermi level) is set to be zero.}
\end{figure}

\begin{figure*}
\centering
\includegraphics[width=0.95\linewidth] {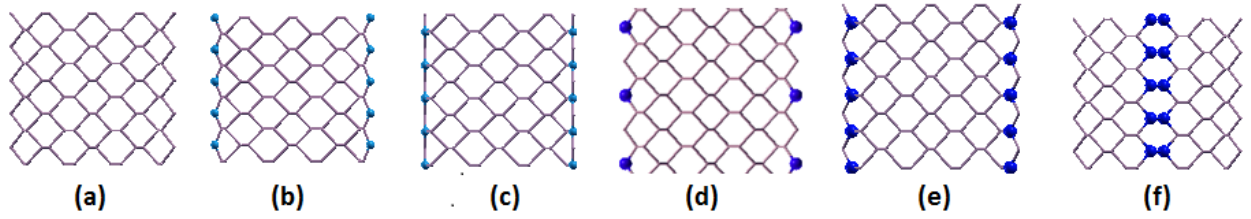}
\includegraphics[width=1\linewidth] {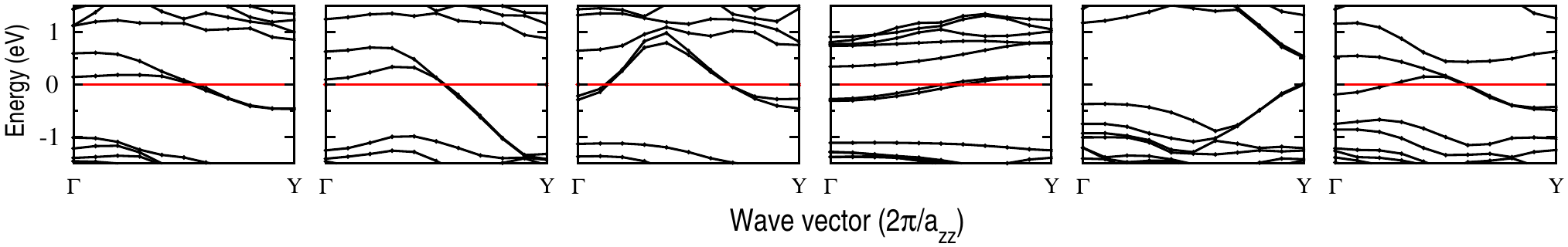}
\caption{\label{pnrStruct} (Color online) Top row: The top view of the fully optimized structure of bare zPNR (a), two types of N-doped zPNRs (b) and (c), and three types of Si-doped zPNRs (d), (e), and (f). The cyan spheres show nitrogen dopant atoms, and the blue (larger) ones are Si dopant atoms. Bottom row: Electronic band structures of the related configurations; in the metallic (semiconductor) systems, the Fermi level (top of the valence band) is set to be zero.}
\end{figure*}

Figure \ref{2dStruct} shows the top view of the pure phosphorene structure, as well as N, Si, and Si-Si doped phosphorene systems. The dopant concentrations in all doped systems are equal. The band structures of given configurations are presented in the lower row of this figure. In phosphorene, each phosphorus atom from the fifth group of the  periodic table, covalently bonds with three other P atoms, creating an $sp^3$ hybridization and forming a semiconductor puckered honeycomb structure with a DFT band gap around $1$ eV. Nitrogen is isovalent with phosphorus, but it has a higher electronegativity, smaller size, and more localized states, which cause a slight increase in the band gap of the N-doped system \cite{gap}. Also, due to its smaller size, it distorts the doped phosphorene structure and as Fig. \ref{2dStruct}(b) indicates, it removes some symmetries and degeneracies of the pristine system. However, a strong similarity exists between N-doped and pure phosphorene band structures.
Si, with four valence electrons, is used for phosphorene p-type doping. It causes the presenting of unpaired electrons in the system, and consequently, it shows metallic behavior. Figure \ref{2dStruct}(d) displays another decoration of the Si-doped phosphorene system, where a pair of P atoms is substituted by a pair of  Si atoms. In this structure, the unpaired electrons are removed by double bonding between Si atoms; as a result, a direct band gap opens at the $X$ symmetry point. However, the band structure of the system is very different from that of the pure phosphorene. Its band gap is close to the gap of the pure system, but it has been shifted from the $\Gamma$ point (in the pure system) to the $X$ point. Due to the dimerization of the unpaired electrons, the Si-doped system shown in Fig. \ref{2dStruct}(d) is energetically more favorable than the structure shown in Fig. \ref{2dStruct}(c).

The formation energy, which shows the energy needed to create the doped system, is calculated as
\be
\label{Eform}
E_{\rm F}=E_{tot}^{(n-m)P+mD}-E_{tot}^{nP}+m\mu_P-m\mu_D
\ee
where $n$ and $m$ indicate the number of P and dopant atoms in the system, $E_{tot}^{(n-m)P+mD}$ ($E_{tot}^{nP}$) denotes the total energy of the system after (before) doping, $\mu_{P}$ is the total energy per phosphorus atom in bulk black P, and finally, $\mu_D$, where $D$ refers to N or Si, is the total energy per N or Si atom, which is respectively obtained by calculating the energy per atom in the N$_2$ molecule or diamond silicon crystal. The formation energies of the doped systems presented in Fig. \ref{2dStruct}, from panels (b) to (d), are respectively 0.73, 0.82, and 0.67 eV per impurity atom. These energies are on the same order as or even less than those of the well-known n- or p-type dopant atoms in silicon structures \cite{silicon}. In addition, we calculate the formation energies of the doped systems regarding three other reference states of P and Si atoms; the results are reported in Ref. \cite{FE}.

In Fig. \ref{pnrStruct}, we present the top view of the bare zPNR system, and five different configurations of N- and Si-doped systems together with their electronic band structures. The widths of all displayed nanoribbons are the same and the dopant concentrations in all doped zPNRs are equal, except for the Si doping concentration in the structure shown in Fig. \ref{pnrStruct}(d), which is half that of the others. Owing to the unpaired electrons of the edge P atoms, the bare zPNR system is metal and midgap states appear around the Fermi level. By doubling the zPNR lattice parameter and halving its Brillouin zone, the degeneracy of the half filled band at the Fermi level is broken and a small band gap ($\sim 10$~meV \cite{pnrtran3}) appears at the Fermi level which is induced by the weak coupling between the unpaired edge electrons. This effect is also known as the Peierls transition \cite{peierls}.

In Figs. \ref{pnrStruct}(b) and \ref{pnrStruct}(c), two different N-doped nanoribbons are displayed. In both structures, dopant atoms are placed at the zigzag edges. Both of these structures are metals and their band structures are comparable to that of the pure system; however, the band structure of the configuration presented in Fig. \ref{pnrStruct}(c) shows more changes due to its stronger distortion.

Figures \ref{pnrStruct}(d), \ref{pnrStruct}(e), and \ref{pnrStruct}(f) show the configurations and the band structures of three different Si-doped zPNR systems. In comparison with the N-doped zPNRs, the lattice distortions of Si-doped systems are negligible but the electronic structures of these systems are very different from the band structure of the pure zPNR due to the lack of an electron in the substitution of P with the Si atom. Because of an odd number of electrons at the edges of the structure displayed in Fig. \ref{pnrStruct}(d), this configuration shows a metallic behavior; the charge density around the Fermi level originates from the edge P and Si atoms.  In the configuration shown in Fig. \ref{pnrStruct}(e), both P and Si edge atoms have an unpaired electron where through dimerization of these electrons, this structure would be a semiconductor. The last configuration presented in Fig. \ref{pnrStruct}(f), has the same dopant concentration as the structure shown in Fig. \ref{pnrStruct}(e), but this structure is metal since its edges are similar to the pure zPNR system. Furthermore, its band structure close to the Fermi level is comparable to that of the pure system. This study reveals the influence of the doping position on the electronic properties of the system.

The calculated formation energies (using Eq. \ref{Eform}) of the doped structures presented in Fig. \ref{pnrStruct} from panels (b) to (f) are respectively 0.62, 0.72, 0.77, 0.22, and 0.67 eVs (see Ref. \cite{FE}, too). The positive $E_{\rm F}$ shows that the doping process is endothermic which is a typical character of the doped systems \cite{silicon}.
The results indicate that for Si dopant, the configuration shown in Fig. \ref{pnrStruct}(e) has a greater chance to be formed due to the dimerization of the edge electrons. However in what follows, we are considering different arrangements of dopant atoms to explore and give a clear view of the doping effect on transport properties of zPNR devices.

\subsection{Transport in zPNRs: Equilibrium conditions}\label{sec:eqTran}

\begin{figure}
\centering
\includegraphics[width=1.0\linewidth] {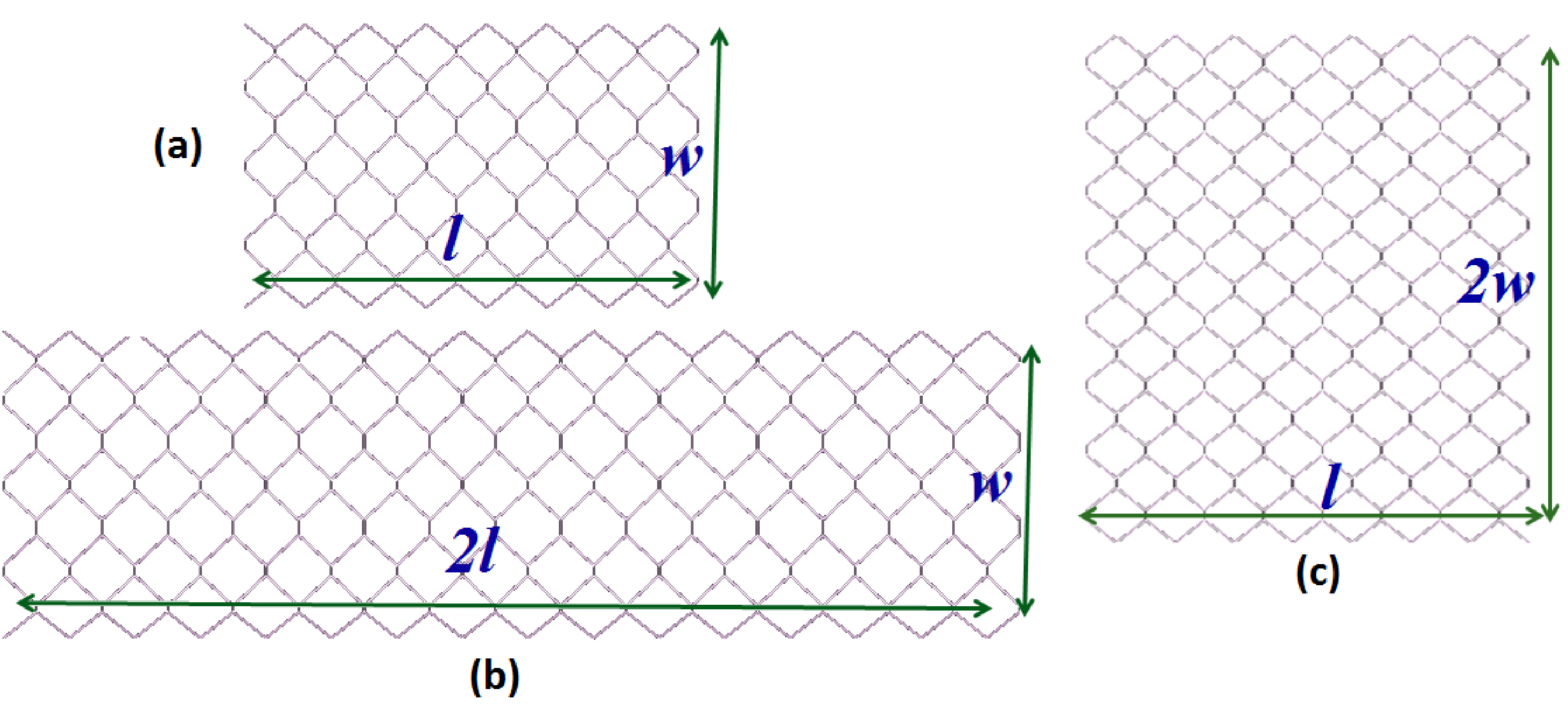}
\includegraphics[width=1.0\linewidth] {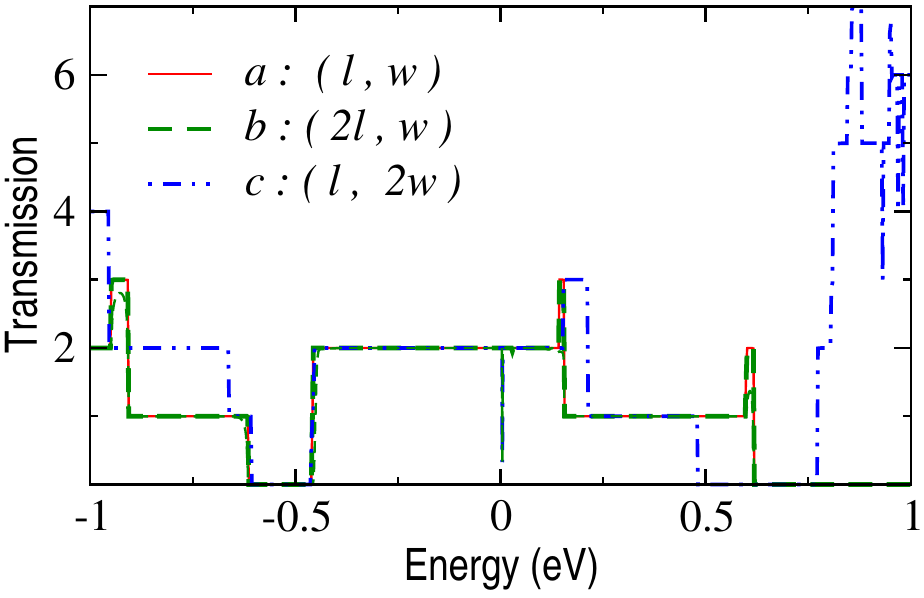}
\caption{\label{Pure} (Color online) Top panels: Top view of the scattering regions of three different sizes of bare zPNR systems. $l=3$ and $w=1.5$ nm denote the length and width of the scattering region of the reference zPNR device shown in (a), and also in Fig. \ref{device}.
Bottom panel: Transmission spectra of the corresponding structures. The solid line presents the transmission profile of the reference zPNR device in the equilibrium conditions (no bias). The dashed (dash-dotted) line shows the transmission spectrum of the zPNR device with two times the length (width).}
\end{figure}

\begin{figure}
\centering
\includegraphics[width=1\linewidth] {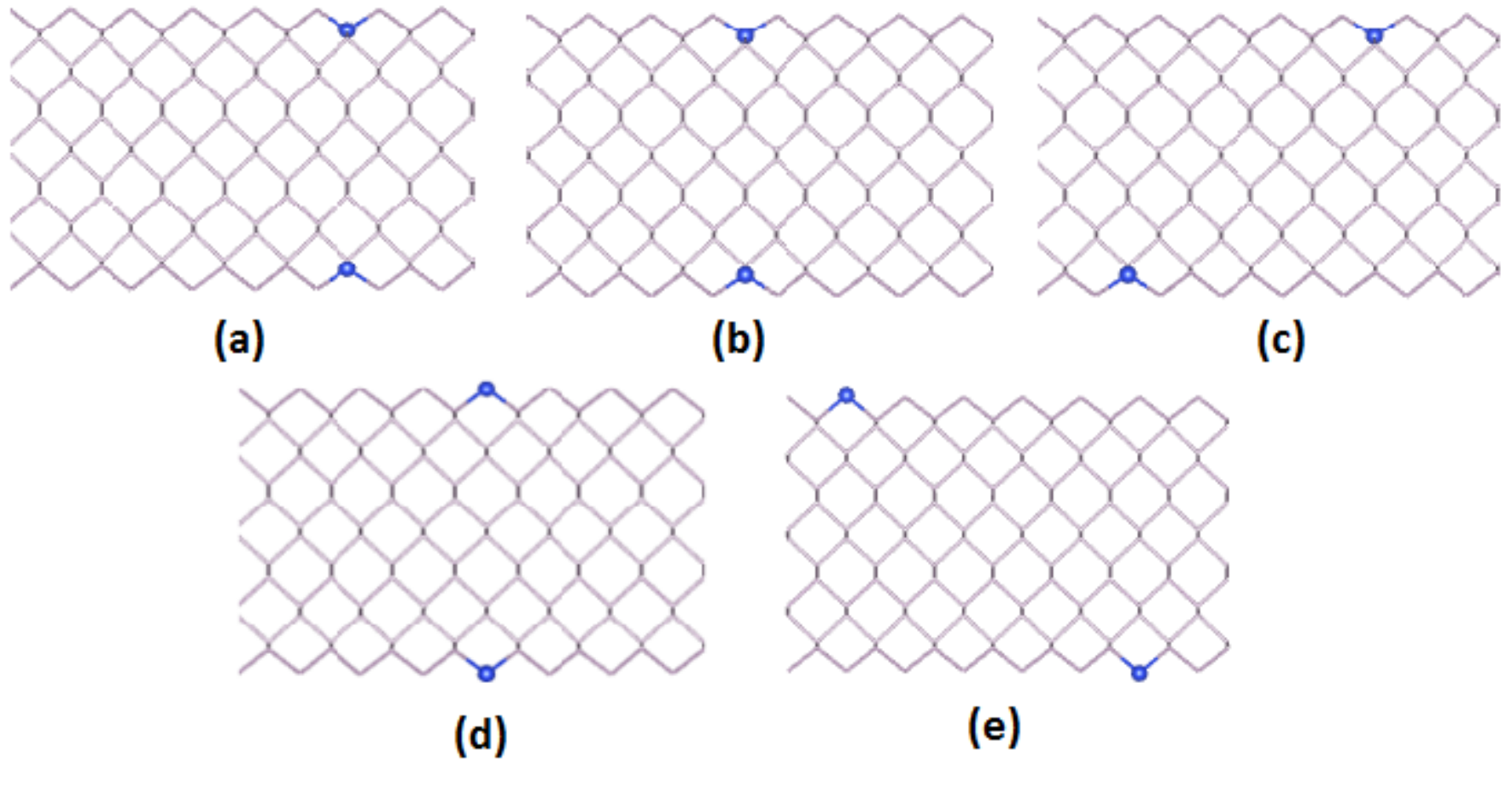}
\includegraphics[width=1\linewidth] {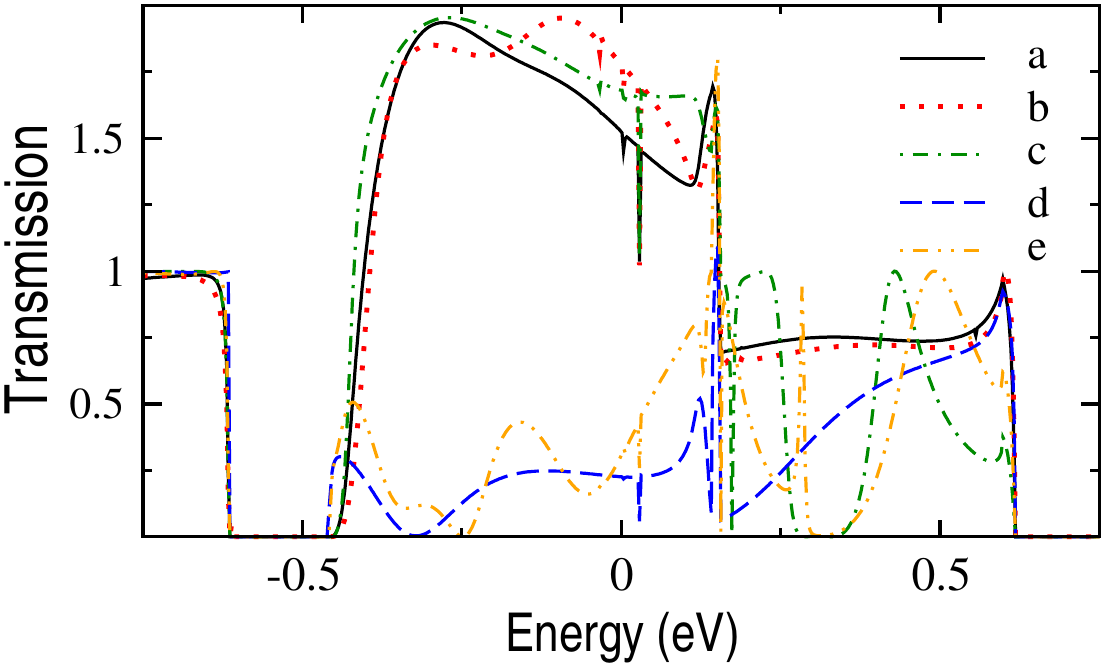}
\caption{\label{SnDop} (Color online) Top panels: Top view of the scattering regions of zPNR devices with single $Si$ dopant atom at each edge. Bottom panel: Transmission spectra of the corresponding doped zPNRs at zero bias, $T(E,~0~V)$. }
\end{figure}

\begin{figure}
\centering
\includegraphics[width=0.97\linewidth] {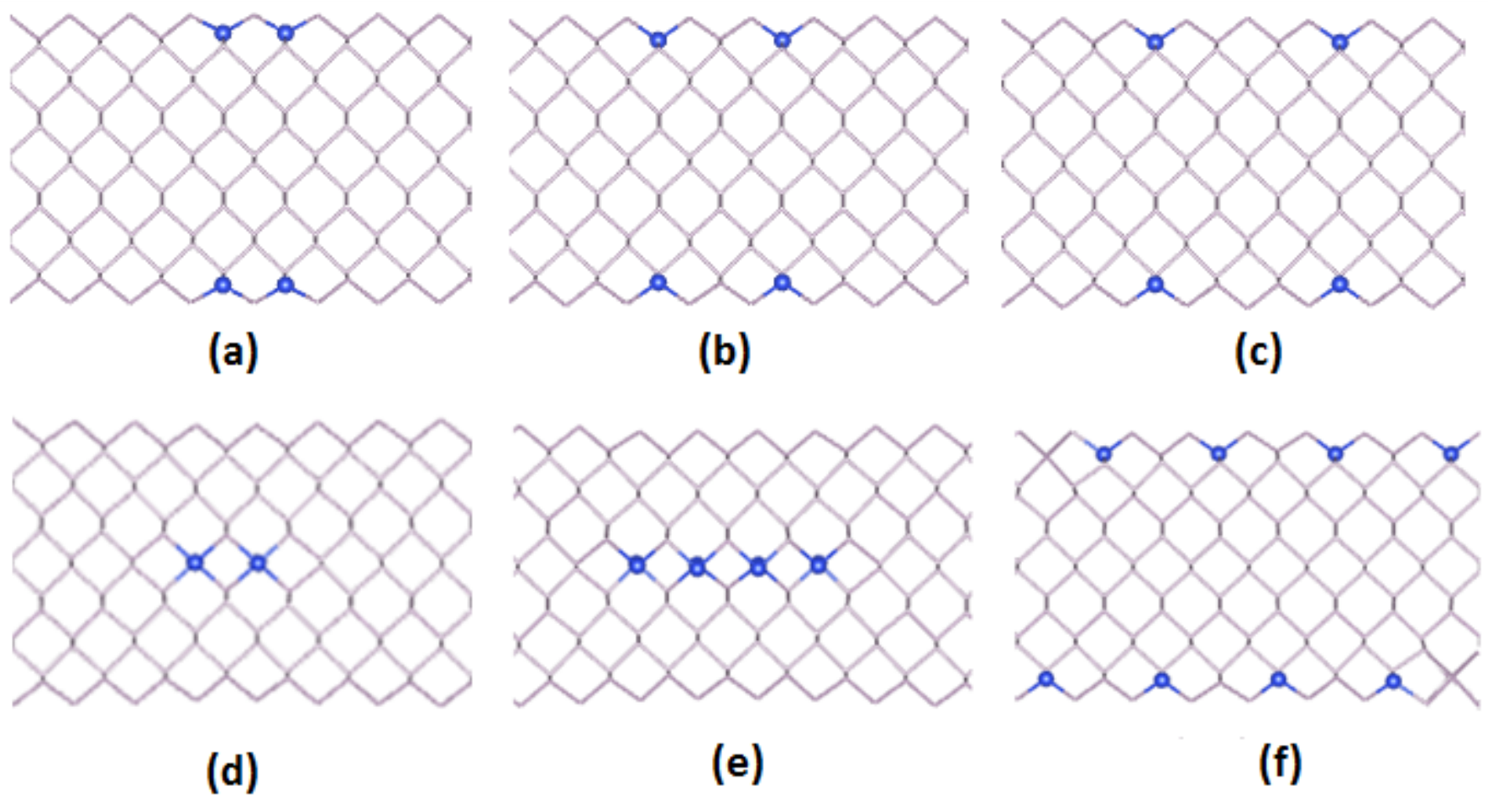}
\includegraphics[width=1\linewidth] {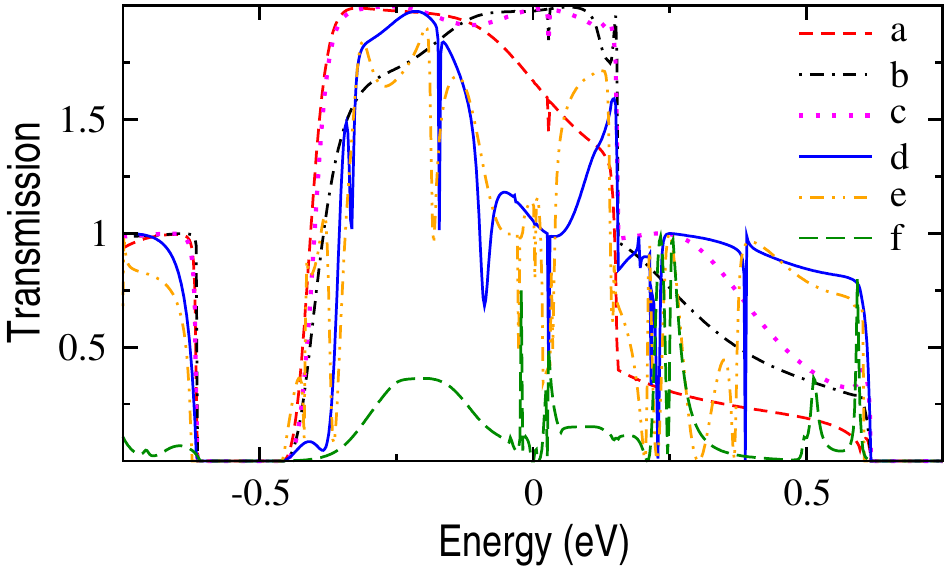}
\caption{\label{DbDop} (Color online) The upper panels show the top view of the scattering region of six different configurations of $Si$ dopant atoms in the zPNR device. Transmission spectra of the corresponding configurations are presented in the lower panel. It shows changes of transmission probability as a function of distance between Si atoms, the dopant concentrations, and the place of the $Si$ impurity with respect to the edges.}
\end{figure}

In this section, we present the results of transport calculations in pure and Si-doped zPNRs under equilibrium conditions. Here, the equilibrium means that the applied bias over the system is zero and both the electrodes have the same chemical potentials. The transmission spectra calculated at equilibrium conditions cannot be used to predict the full features of the transport properties of the systems with individual quantized levels \cite{datta1,datta2} since the linear response of the transport is valid when the applied bias voltage is very small compared to the system energy level broadening and its thermal energy  \cite{datta1,datta2}. Therefore, as we mentioned in Sec. \ref{sec:method}, to explore zPNR devices, we need to perform the more expensive and complicated self-consistent calculations, including solving the Poisson equation under the applied bias \cite{datta1}. However, this study could reveal some features of the charge transport around the zero bias. 

Figure \ref{Pure} shows the calculated transmission spectra of three different sizes of a pure zPNR device. The central regions of the related structures are shown in the upper panel of this figure. $l$ and $w$ refer to the length and width of our reference device, displayed in Figs. \ref{device} and \ref{Pure}(a). According to Fig. \ref{Pure}, the transmission spectra of devices with the same width but different lengths of $l$ and $2l$ are exactly equal. This is an indication of the ballistic transport where the carriers propagate without any scattering, unlike the Ohmic transport that resistance is in proportion to the device length. This figure also gives a comparison between the reference device and a twice wider device; their transmission spectra around the Fermi level are same as each other. This suggests that under low biases, transport in zPNRs is carried out by the carriers localized near the ribbon edges. Additionally, the quantized transmission plateaus are further evidence of the ballistic transport regime in the pure zPNR system. The transmission value at this regime is equal to the total number of the transmitting channels in the corresponding energy. The sharp antiresonance minimum at the Fermi level of the transmission spectra in Fig. \ref{Pure} arises from the small band gap of the zPNR system discussed earlier in Sec. \ref{sec:struct}.

In Fig. \ref{SnDop}, we present the scattering region of five different Si-doped zPNR structures with a single defect at each edge.  The electrodes are pure zPNR systems. Our aim of this study is exploring the impact of impurity position on the transport properties of the system. In the first three (last two) structures, the impurity atoms are placed in the same symmetry points of the lattice, but at different distances against to the electrodes. Regarding the transmission spectra appearing in the lower panel of Fig. \ref{SnDop},  the transmissions are almost suppressed in the configurations shown in Figs. \ref{SnDop}(d) and \ref{SnDop}(f) due to the strong scattering or non availability of the appropriate states in the scattering region.
These two configurations are energetically less favorable than the other presented structures, since Si atoms lose a couple of their bonds and thus they have a high potential to trap electrons. Also, as will be shown in the following section, at these two configurations, impurity blocks two of the main transport channels.
The transmission behaviors of other structures are not the same, but they are close to each other. Also, the ballistic transport no longer exists in the doped systems due to the appearance of the scattering centers.

Figure \ref{DbDop} displays the central region of six different Si-doped zPNR configurations. Our aim of this study is threefold. The first one is finding the impact of the distance between atomic defects on transport, the next is showing the effect of impurity positions with respect to the edges on transport, and the last one is displaying the influence of the number of impurities on transport. The transmission profiles of the given structures are exhibited in the lower panel of Fig. \ref{DbDop}. There are three configurations with double atomic defects on each edge. The configuration shown in Fig. \ref{DbDop}(a), with a closer distance between dopant atoms, shows the smaller transmission around the Fermi level. The transmission of two other configurations in the vicinity of the Fermi energy is almost similar. Changes of the energy levels of the structures presented in Figs. \ref{DbDop}(b) and \ref{DbDop}(c)  are in a way that their transmission spectra  around the Fermi level are stronger than the transmission probability of the single Si-dopant structure displayed in Fig. \ref{SnDop}.
By shifting the place of Si dopant atoms from the edges to the central part of the ribbon, the transmission around the Fermi energy is reduced; because, as shown in Fig. \ref{pnrStruct}(f), this kind of defect could manipulate the states close to the Fermi level, but its transmission is robust against increasing the number of the central dopant atoms. However, through increasing the number of the scattering centers in the edges, the conducting channels are almost blocked and transmission is strongly reduced.

\subsection{Transport in zPNRs: Finite bias results}\label{sec:neTran}

In this section, we investigate the conductance properties of the pure zPNR as well as eight different configurations of N- and Si-doped zPNR devices, under an applied bias voltage using fully self-consistent electronic transport calculations on the basis of the NEGF formalism, introduced in Sec. \ref{sec:method}.

The $I-V$ curve characteristic of pure zPNR devices has been reported previously \cite{pnrtran2,pnrtran3,pnrtran4}, however, all features of transport have not been studied.
For the sake of completeness, we comprehensively calculate the voltage drop, transmission pathways, atomic current, and effect of the N and Si doping on zPNR systems in the out-of-equilibrium conditions. Moreover, the size of our considered device is different from theirs; therefore, one can explore the influence of system size on the transport properties.

From the physical point of view, a typical acceptable applied bias over a nanostructure is only a few volts; in order to be less than the breakdown voltage \cite{datta1}. Thus, our focus is on the biases in the range of $0-1.7$ ($0-1$) in the pure (doped) zPNR structure(s).

\subsubsection{Pure system}

\begin{figure}
\centering
\includegraphics[width=1.\linewidth] {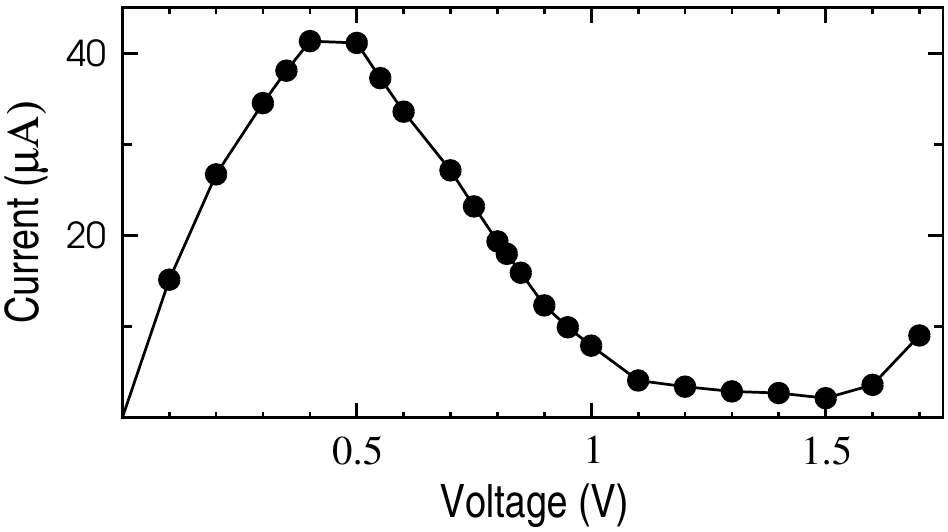}
\caption{\label{ivPure} (Color online) $I-V$ characteristic of the pure zPNR device.}
\end{figure}

Figure \ref{ivPure} shows the non linear $I-V$ curve of the pure zPNR device in the bias range between $0$ and $1.7$ V. As the figure indicates, the current starts increasing linearly as a function of the applied bias until the bias (total current) equal to $0.4$ V ($\sim 40~\mu A$); then a small plateau appears in the curve in the bias area of $0.4-0.5$ V and afterwards, the current decreases by increasing the applied bias in the bias range between $0.5-1.1$ V. This feature is known as the negative differential resistance (NDR) phenomenon. Another plateau occurs in the valley area of the $I-V$ curve in the bias range of $\sim~1.1-1.6$ V, and then the current increases again under the applied bias.
The peak-to-valley ratio (PVR), which is determined as the ratio of the current at the resonant tunneling peak energy to that at the valley, is an important parameter and it is obtained $\sim 25$ for this system. The obvious low bias NDR is important to reduce a power consumption of the electronic devices. Also the plateaus appearing at the peak and valley are useful characters for practical applications.

Stemming from our numerical results shown in Fig. \ref{ivPure}, the current intensity or the magnitude of the electric current is on the order of several tens of micro amperes, which is in the same range as the high current intensity calculated in graphene nanoribbons \cite{gnrTran} or the bare zBN nanoribbons \cite{hBN-tran}; it is several orders of magnitude higher than the current of H-terminated zPNR \cite{pnrtran2} and bare armchair PNR \cite{pnrtran3} in the same voltages. The PVR value in this system is comparable to the reported values in resonant tunneling diodes composed of group $III-V$ compounds, such as GaAs/InGaAs/AlAs with a PVR  of $\sim 5$ \cite{rtd}, or a PVR value of  $50-200$ in defected aGNRs \cite{agnr}, or $5.5$ in the z-MoS$_2$ nanoribbon \cite{MoS2-tran}, but very small in comparison with the observed PVR value in CdSe quantum dots ($\sim 1000$) \cite{cdse}.

The overall picture of the $I-V$  characteristic in our system is very similar to that of the zPNR device studied in Ref. \cite{pnrtran1} which is twice wider than our system, because the dominant part of the current arises from the edges. However, there are some differences in the higher bias voltages due to the small contribution of the central part of the system in the transport at higher biases. As a result, the reported PVR in the twice wider system is equal to $7.4$ which is one-third of ours. This difference originates from the higher valley current of the wider system. Also, the NDR bias window in their work is smaller than ours. In their system at the bias voltage of $0.85$ V, the current starts increasing. Moreover, the plateau around the valley has not been seen in their works (see also~\cite{pnrtran2}).  Consequently, the ribbon width can be used as a way to control and improve the NDR characteristic.

\begin{figure}
\centering
\includegraphics[width=1.\linewidth] {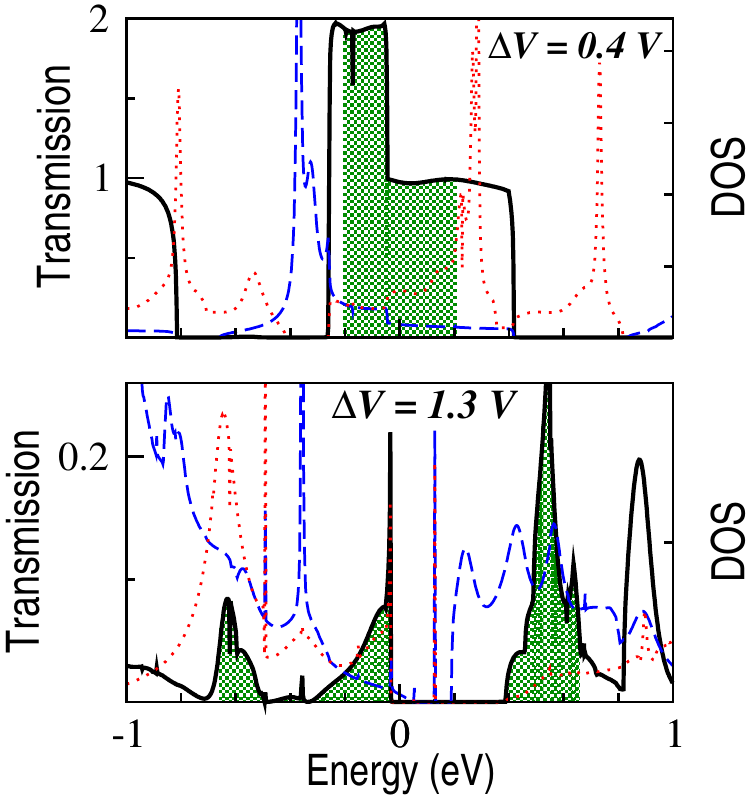}
\caption{\label{dosTran} (Color online) The solid line in the upper (lower) panel  shows the transmission spectrum of the zPNR device under the bias voltage of $0.4$ ($1.3$) V as a function of carrier energy. The dashed (dot-dashed) lines denote  the density of states of the left (right) electrodes as a function of energy under the corresponding biases. The shaded areas contribute to the current intensity.}
\end{figure}

The NDR phenomenon is the characteristic of the resonant tunneling devices \cite{datta1,datta2}. It was observed for the first time in highly doped Ge systems \cite{esaki} leading to the discovery of the electron tunneling effect which was recognized with the Nobel prize in physics in $1973$. The bias voltage by shifting up and down the discrete density of states, could provide the conditions that individual energy levels of the central region are located in the gap area of electrodes and feel two potential barriers at two ends of the scattering region, leading to the resonant tunneling transport and observed NDR behavior.

In order to explain the NDR mechanism in this system, we present the transmission spectra, as well as the density of states of the left and right electrodes in Fig. \ref{dosTran} under the bias voltages of $0.4$ V and $1.3$ V, located in the peak and valley areas of the $I-V$ curve, respectively. The quantized plateaus appearing in the transmission spectrum at the bias of $0.4$ V (see the upper panel of Fig. \ref{dosTran}) indicate that the carriers pass through the central region almost without reflection, pointing out the ballistic transport regime. The value of the quantized transmission in the ballistic regime is equal to the number of the conducting channels available at the given energy under the bias voltage. The shaded area shows the integration of the transmission over the bias window where based on the Landauer formula, Eq. \ref{landaue}, it is proportional to the current intensity. According to the results presented in Fig. \ref{dosTran}, the overlap between the density of states of the left and right electrodes is necessary to have a non zero transmission, but it is not the sufficient condition and the availability of the appropriate energy levels in the central region is also required. The transmission spectrum under the bias voltage of $1.3$ V indicates that high reflection appears in the system which arises from the band alignment at the contacts. Also, there is no overlap between electrode DOSs for the entire region of the higher bias of $1.3$ V. Accordingly, both of these effects, which are coming from shifting the discrete electrode DOSs via the electric bias, result in reducing the transmission and occurrence NDR in the system.

\begin{figure}
\centering
\includegraphics[width=1.\linewidth] {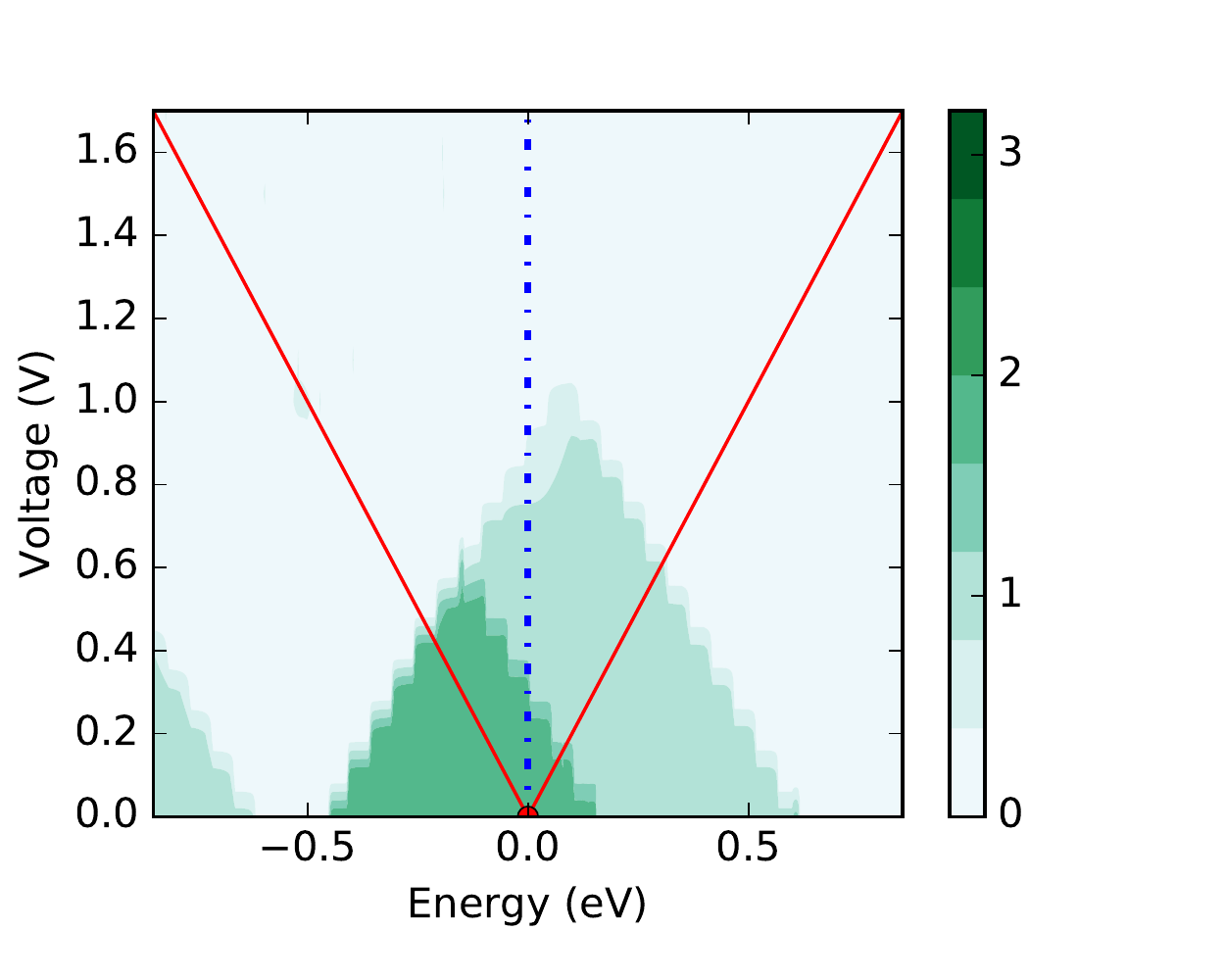}
\caption{\label{tranPure} (Color online) The transmission spectrum of the zPNR system as a function of charge carrier energy and bias voltage. The transmission integration over the bias window, determined by the solid red lines, gives the current intensity under the applied bias voltage. The dashed line shows the Fermi level which is set to be zero.}
\end{figure}

The transmission spectrum of the zPNR system as a function of the carrier energy and bias voltage is illustrated in Fig. \ref{tranPure}. The total current at a given bias voltage is defined by the transmission integration in the bias window expressed by two solid lines. Note that transmission in the negative (positive) energies refers to the conductivity induced by holes (electrons). Therefore, in the bias range between $\sim 0-0.5$ V, the dominant part of the conductance comes from holes. The NDR behavior accompanies the changing of the type of the contributing carriers in the transport from holes to electrons or, in other words, from the valence band to the conduction band.

\begin{figure}
\centering
\includegraphics[width=0.95\linewidth] {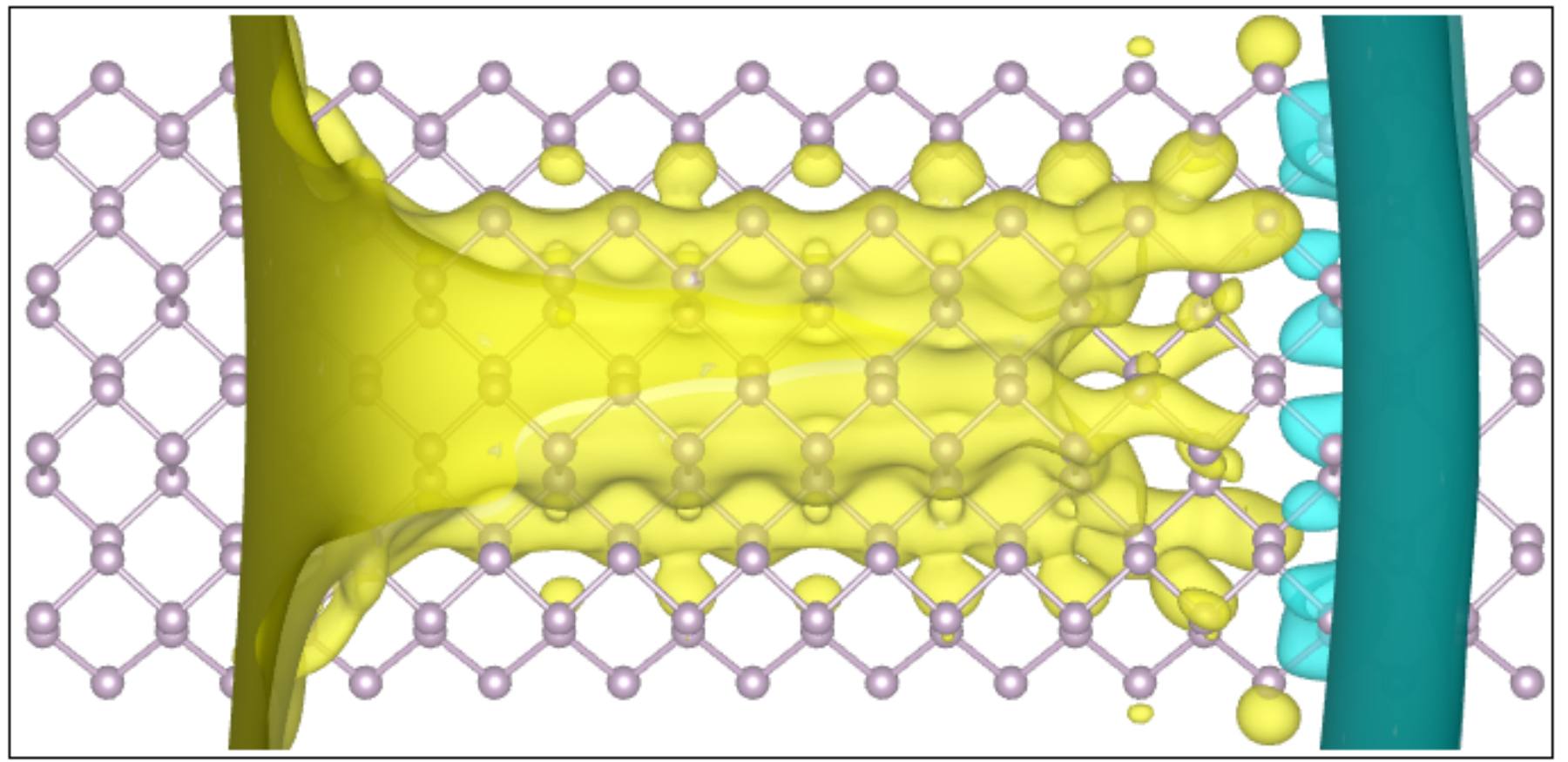}
\includegraphics[width=0.45\linewidth, angle=90] {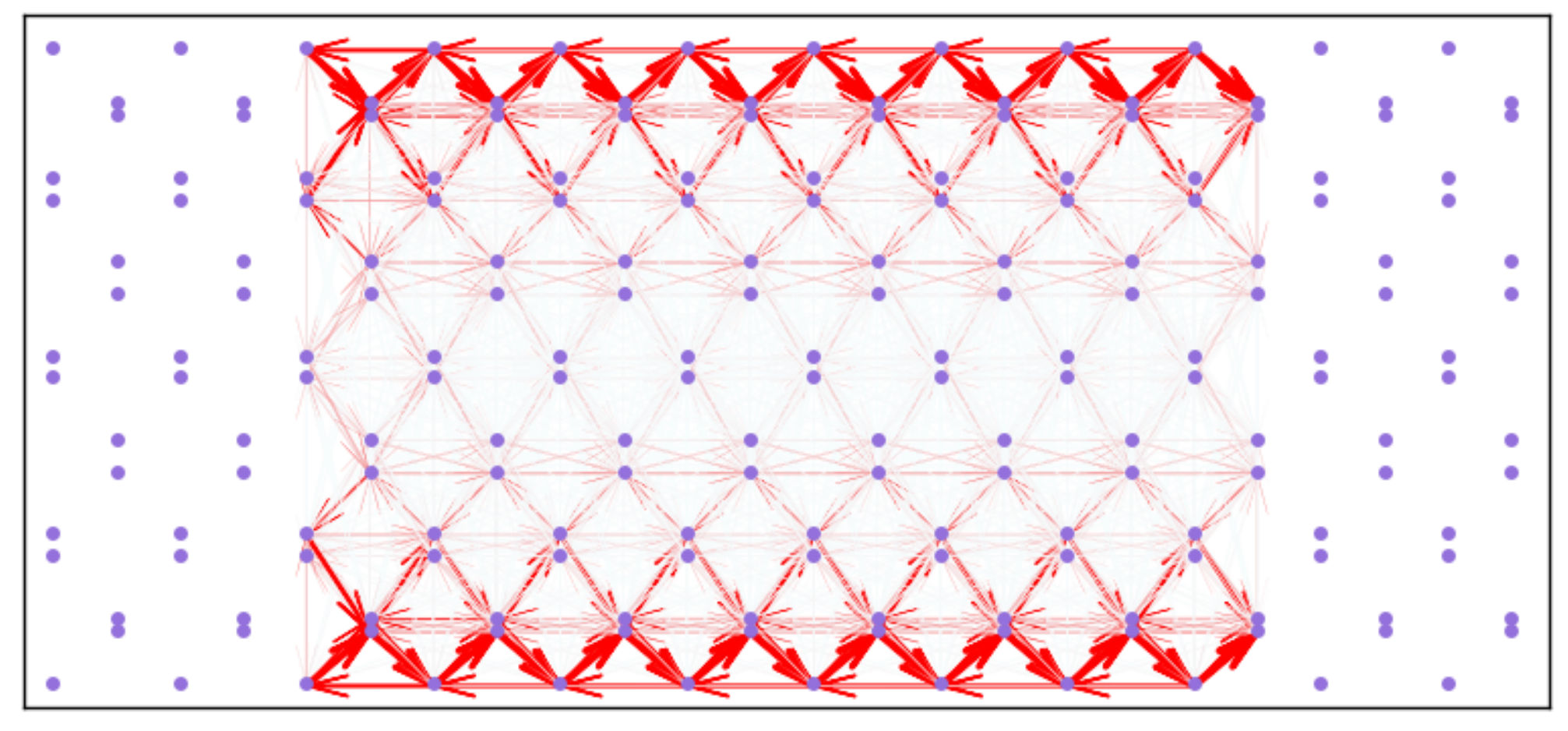}
\caption{\label{vPure} (Color online) Top panel: The calculated voltage drop of the zPNR system across the central scattering region at a bias of $0.4$ V. The surface isovalue is $0.15~eV\AA^{-3}$. The potential is strongly pinned to the left (positive) electrode.
Bottom panel: Transmission pathway at a bias of $0.4$ V, which shows how current flows through the device. The width of the arrows is related to the magnitude of the local current density. The local currents are totaled over the energy bias window.}
\end{figure}

In order to investigate more details of the charge transport mechanism in our device, we calculate the voltage drop and transmission pathway (also known as bond current) of our pure system under the electric bias of $0.4$ V, and the results are displayed in Fig. \ref{vPure}. The voltage drop or potential profile is the difference between the effective potential (here Hartree electrostatic potential) under a bias voltage and in equilibrium conditions. Thus, it visualizes the small potential changes induced by the applied bias. The result presented in the upper panel of Fig. \ref{vPure} indicates that a high voltage drop happens in the central part of the nanoribbon close to the right electrode. Since the voltage drop is a measure of the device resistance \cite{datta1,potDrop}, we conclude that this area has a higher resistance against current flowing compared to the edges. This behavior is confirmed by the transmission pathway displayed in the lower panel of Fig. \ref{vPure} which indicates that the current flowing in the central part of the nanoribbon is negligible and the conductance channels are localized at two edges.

The bond current or transmission pathway \cite{handbook} is an analysis tool which is derived from the splitting of the transmission coefficients into the bond directions and gives a pictorial explanation of the current flow in the system. Note that information about the mathematical equations of this quantity can be found in Refs. \cite{sisl} and \cite{bondi}.  The bond current illustrated in the lower panel of Fig. \ref{vPure} is totaled over the bias window. The directions of the arrows show the  direction of the local current density flow and the widths of the arrows correspond to the magnitude of the current density between the related pair of atoms. Regarding Fig. \ref{vPure}, the transmission arises from two different types of currents, namely the bonding current which carriers pass through the chemical bonds, and the hopping current which comes from the charge hopping between pairs of atoms without any chemical bond connection.  As shown in the lower panel of Fig.  \ref{vPure}, the dominant part of current passes through the chemical bonds at the zigzag edges. Also, a kind of loop current arises from the charge hopping between the next nearest neighbor P atoms of the same plane. This kind of the loop current is similar to the loop current in graphene nanoribbons \cite{gnrTran}. Besides, the loop current has been seen in a benzene ring of defected graphene \cite{bcur1,bcur2} and some kinds of molecular junctions \cite{pway}, but the loop current of these systems arises from the chemical bonds. A small current flowing is additionally observed between P atoms in the upper and lower planes.
Regarding the localization of the transmission channels at the edges, the zPNR device behaves like a truly 1D resonant tunneling device under the bias voltage.
The edge transport is the typical properties of nanoribbon devices  \cite{tran-edge1,tran-edge2,gnrTran}.

\begin{figure}
\centering
\includegraphics[width=1.\linewidth] {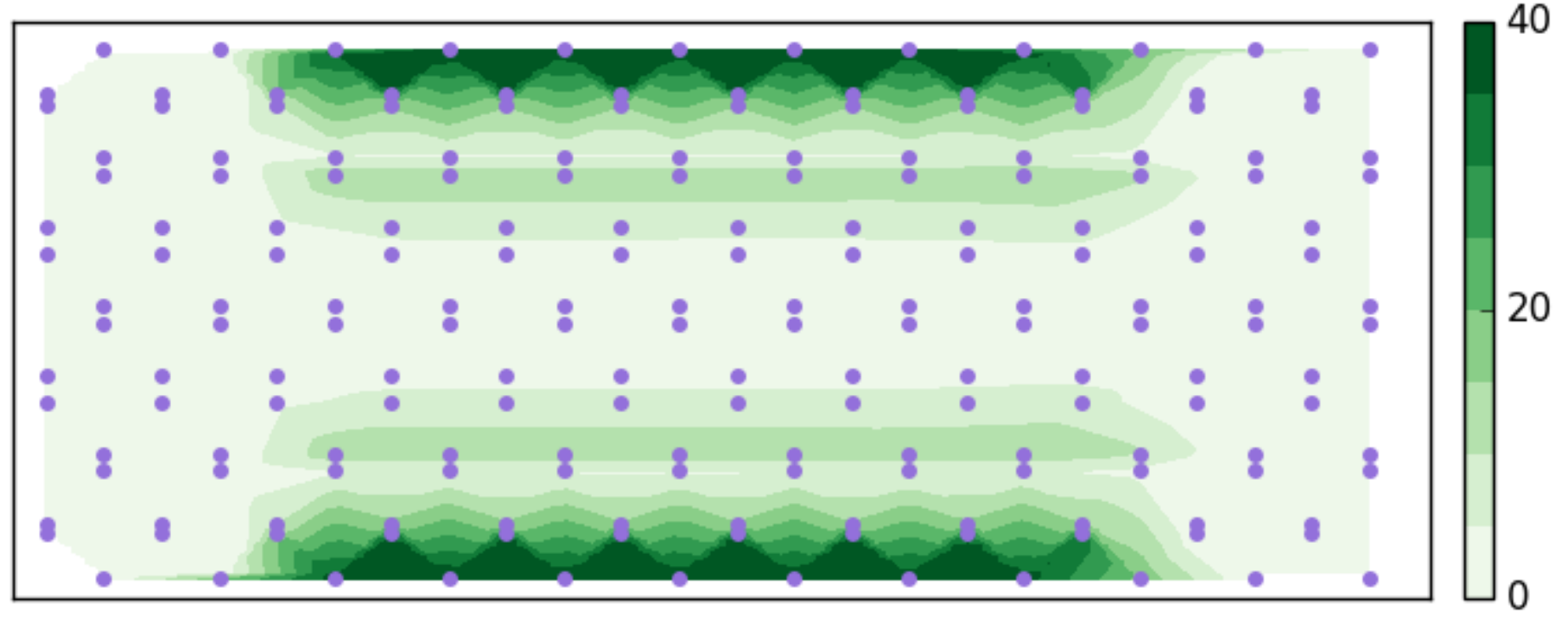}
\caption{\label{AtmIPure} (Color online) The contour plot of the atomic current of the zPNR device under a bias voltage of $0.4$ V. It reveals the local current density of the system. The atomic current values are integrated in the bias window.}
\end{figure}

In Fig. \ref{AtmIPure}, we present another transport property of the zPNR system, named the atomic current \cite{sisl}. The atomic current is a scalar quantity with a single value for each atom of the device.  It thus gives an idea of the atom contribution to the current flowing. Figure \ref{AtmIPure} shows a contour plot of atomic current of the zPNR system under an electric bias of $0.4$ V; the atomic current quantities are totaled over the bias window. Continuing this scalar quantity will be used to give an intuitive picture of the influences of doping in the system.

\subsubsection{Dopant effects}

\begin{figure*}
\centering
\includegraphics[width=1\linewidth] {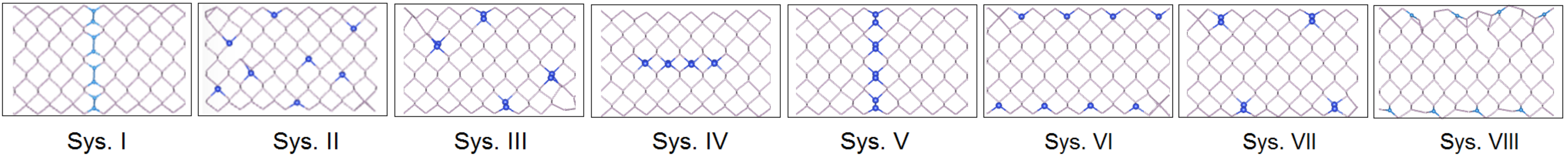}
\caption{\label{struct}(Color online) The optimized structures of the proposed $N$- and $Si$-doped zPNR devices. The cyan spheres in the first and last systems, show $N$ atoms; $Si$ atoms are in blue (the larger spheres).}
\end{figure*}

\begin{figure}
\centering
\includegraphics[width=1\linewidth] {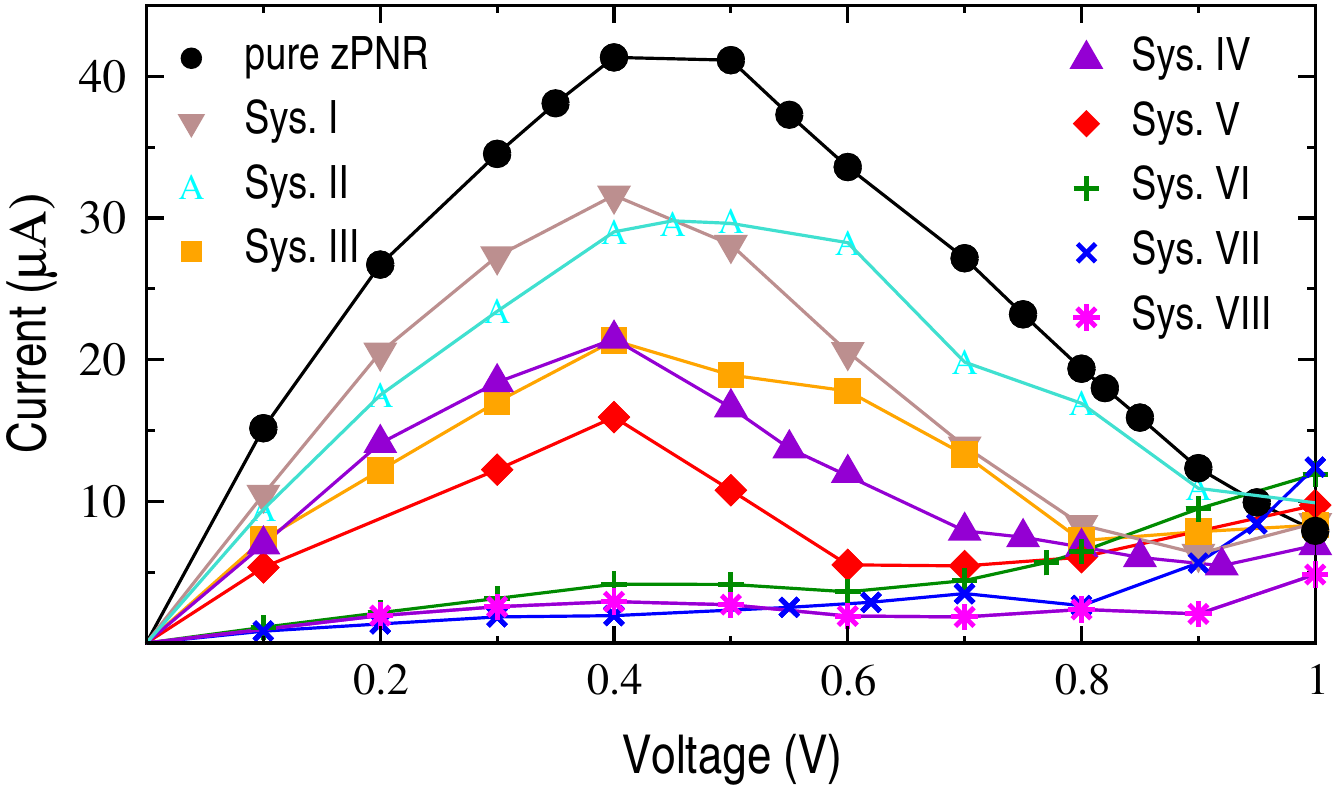}
\caption{\label{IV}  (Color online) $I-V$ characteristics of the doped zPNR structures shown in Fig. \ref{struct}. The current is strongly suppressed by the edge disorder. The $I-V$ curve of the pure system has been plotted as well.}
\end{figure}

\begin{figure}
\centering
\includegraphics[width=1\linewidth] {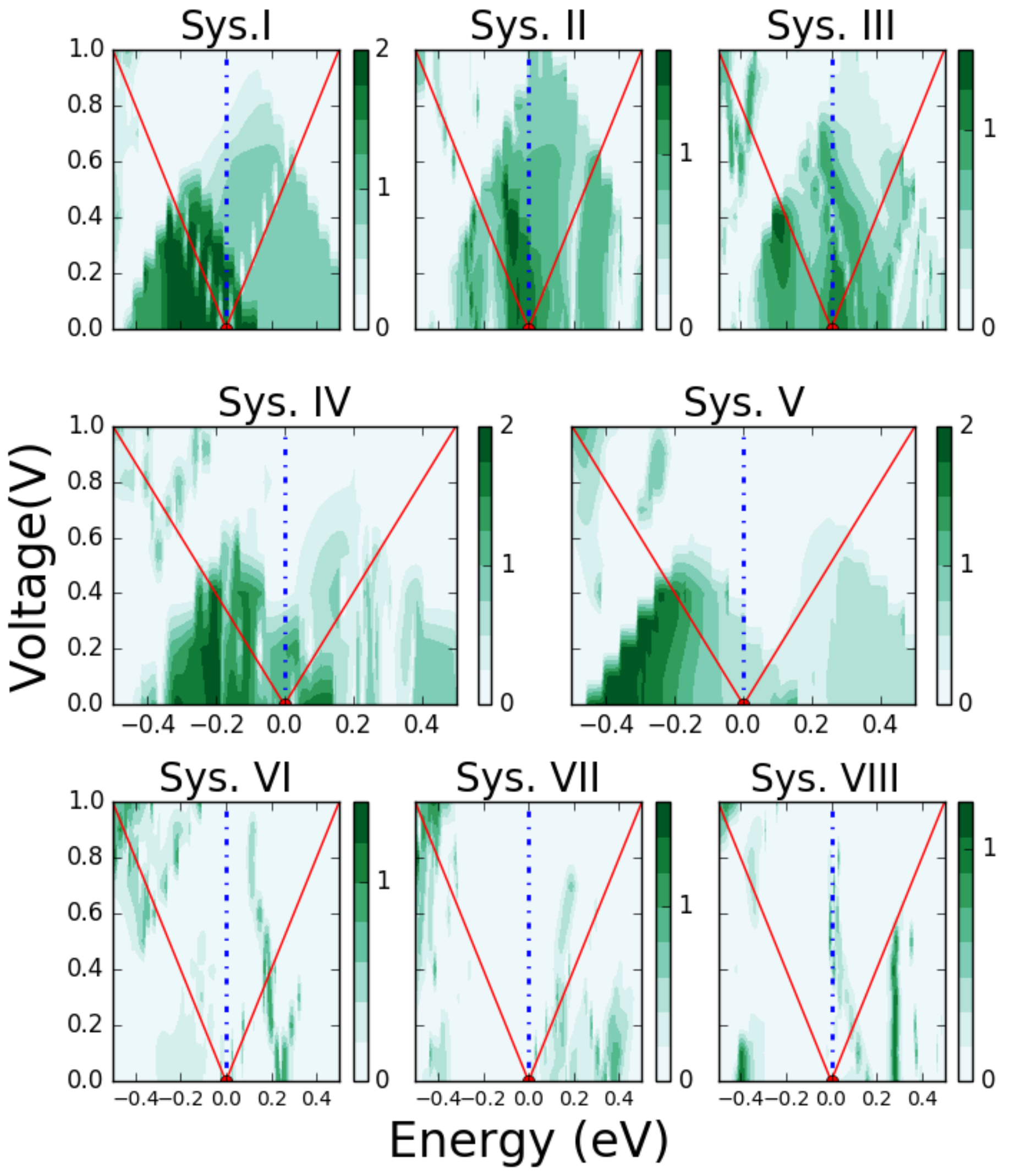}
\caption{\label{tran} (Color online) Transmission spectra as functions of the carrier energy and applied bias for the doped zPNR systems where their corresponding geometries are shown in Fig. \ref{struct}.}
\end{figure}

\begin{figure}
\centering
\includegraphics[width=1\linewidth] {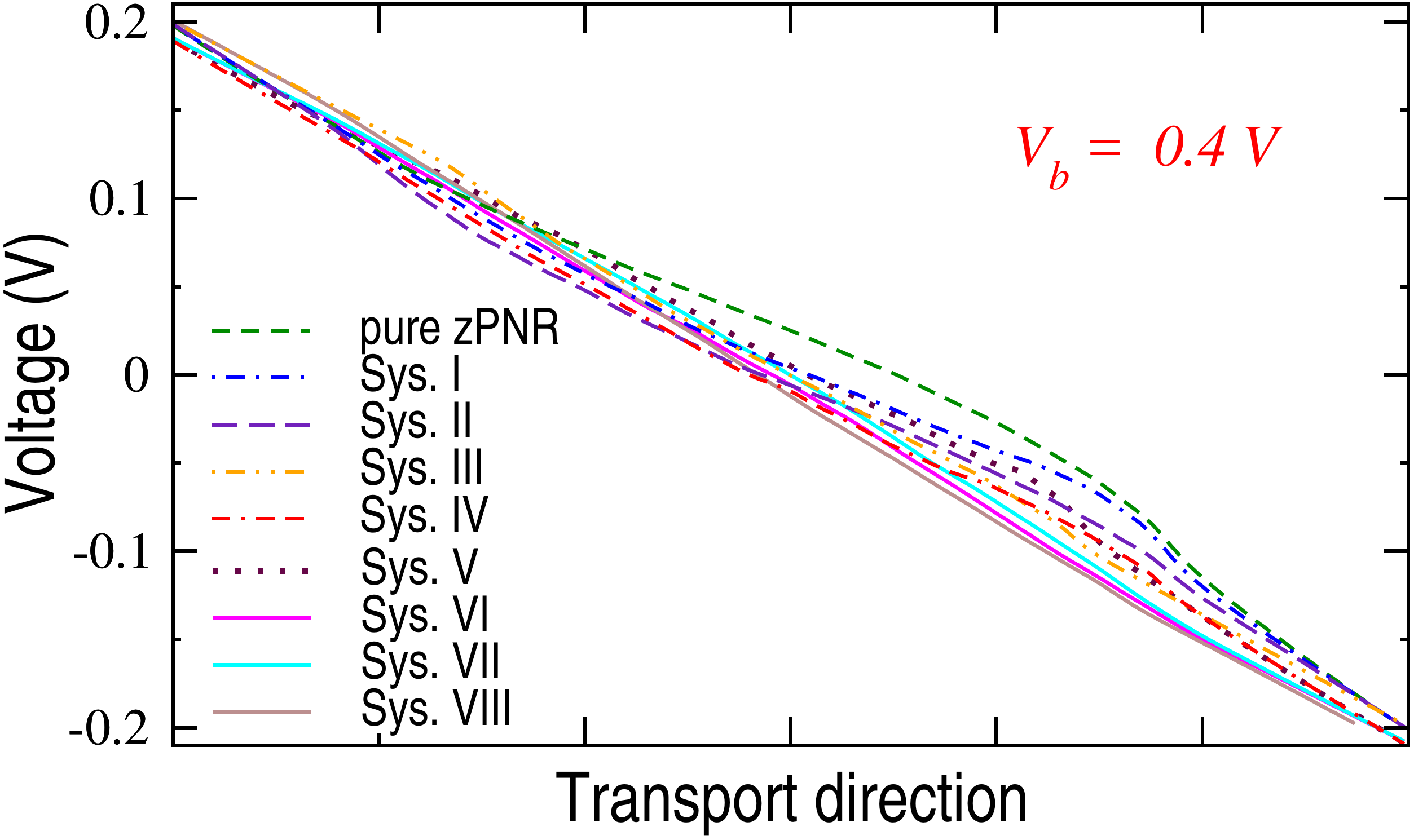}
\caption{\label{vDrop} (Color online) The self-consistent voltage drop along the transport direction for the doped and pure zPNR devices at an external bias of $0.4$ V. It has been averaged in the two other directions perpendicular to transport.}
\end{figure}

\begin{figure}
\centering
\includegraphics[width=1.\linewidth] {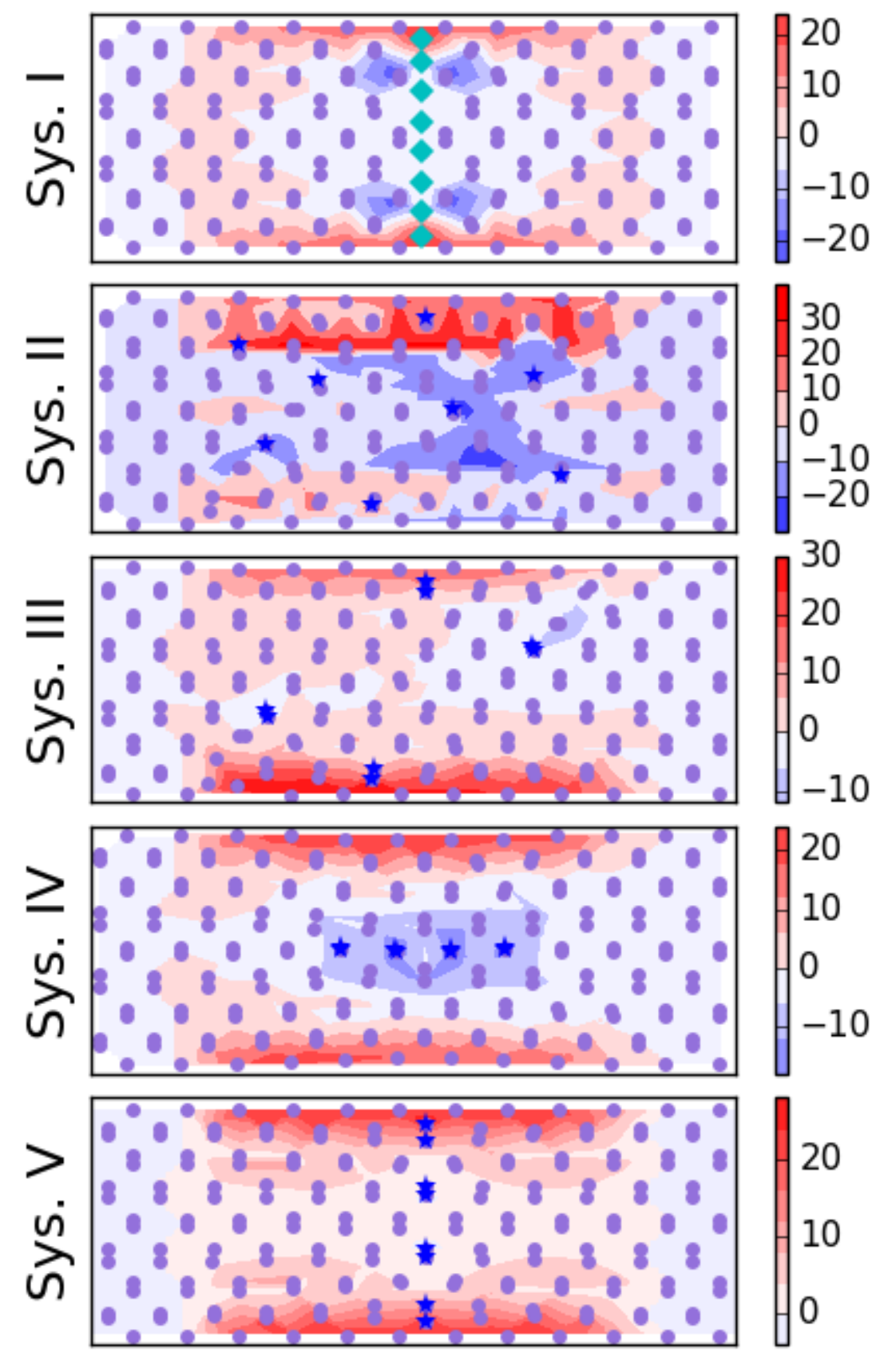}
\caption{\label{AtmI} (Color online)
 The subtraction of the atomic currents of those doped systems which show higher conductivity as well as the NDR characteristic from that of the pure system (presented in Fig. \ref{AtmIPure}).  The blue stars (cyan diamonds) show Si (N) dopant atoms in the zPNR background. The atomic current of the corresponding doped systems is illustrated in the Appendix.}
\end{figure}

\begin{figure*}
	\centering
	\includegraphics[width=1.\linewidth] {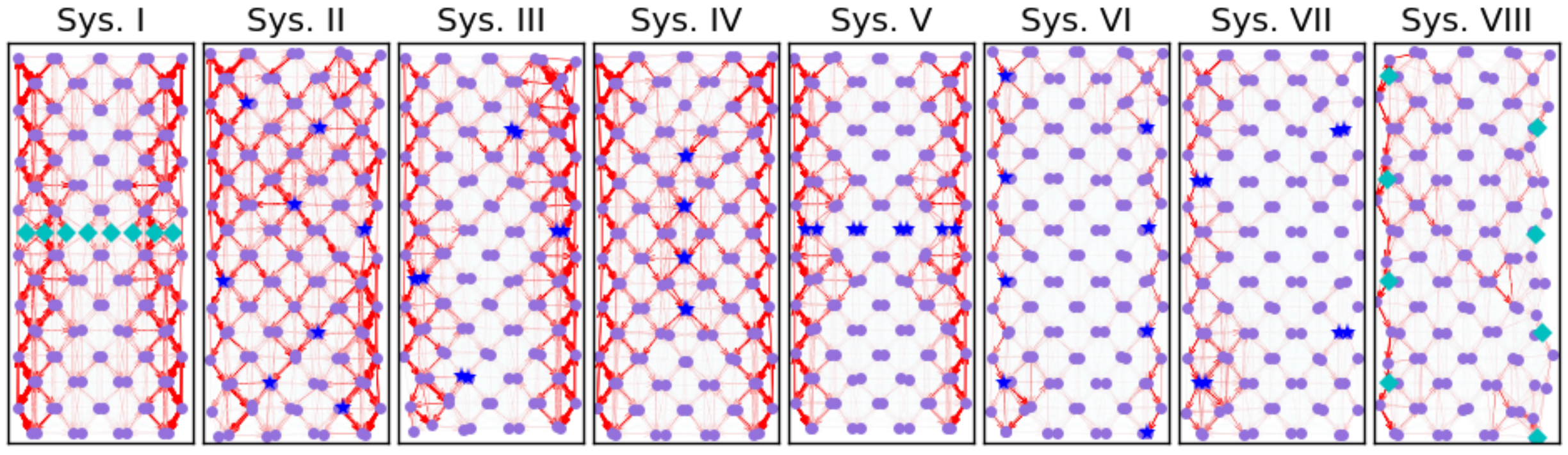}
	\caption{\label{BndI}  (Color online)
		Visualization of the transmission pathways (or bond current) through the scattering region of the studied doped systems. The blue stars (cyan diamonds) present $Si$ ($N$) dopant atoms in zPNR background. The $Si$ atoms at the central part of the nanoribbons behave like scattering centers, while dopant atoms at the edges are like a current blocker. Regarding the positions of dopant atoms, the diffusive transport regime as well as the localized regime are observed.}
\end{figure*}

Now we turn our attention to the transport properties of the doped zPNR devices under an applied bias. Figure \ref{struct} shows the central part of our studied systems where we consider eight different decorations of N- and Si-doped zPNR devices. The number of the dopant atoms is equal in all structures and the size of devices is the same as our pure zPNR system presented in the previous part. The atomic sizes of P and Si are close to each other, but the smaller size of N atoms induces strong distortion in the N-doped systems. The electrode parts, which are not presented in the figure, are made of pure zPNR.

The $I-V$ characteristics of the doped and pure zPNR devices are displayed in Fig. \ref{IV} for the bias range of $0-1$ V. Due to the left-right symmetry of the proposed structures, we just consider the positive biases. The dopant atoms increase the device resistance and reduce the current intensity in the presented bias values. This reduction in the edge defected systems (the last three configurations shown in Fig. \ref{struct}) is stronger and the current intensity decreases by an order of magnitude, and furthermore the NDR characteristic is completely suppressed. The N- and Si-doped configurations displayed in systems I and II show the highest conductivity among the studied doped systems. The loss of the current in the N-doped system compared to the pure device is $\sim 20\%$ (around the peak area), while for the same configuration of Si dopant atoms (system V), the current loss is $\sim 70\%$. This reveals the effect of conserving the $sp^3$ hybridization in the transport properties of the system. In the second system, the current reduction around the peak bias is $\sim 20\%$  but in the higher biases, its current intensity is close to the current of the pure system.  A strong current reduction ($\sim 50\%$) is also observed in system IV without any edge defects.
The first five configurations shown in Fig. \ref{struct} exhibit the NDR characteristic. Also PVR parameters are reduced from $25$ in the pure system to $5$, $3$, $3$, $4$, and $2.9$ in systems I--V, respectively.

The transmission spectra of the doped systems as functions of the carrier energy and applied bias are illustrated in Fig. \ref{tran}. The solid red lines in the figure correspond to the bias window. The lowest transmissions occur at the edge defected systems. In the other structures except systems II and III, the dominant part of the conductance comes from the hole contribution, especially at the lower biases. In systems II and  III, both electrons and holes contribute in the transport and consequently, the conductance reduces more slowly at the NDR region.

Figure \ref{vDrop} shows how voltage drops through the central part of the pure and doped systems under the bias voltage of $0.4$ V. The voltage drop is plotted along the transport direction and it is averaged in the perpendicular directions. As the figure indicates, the voltage drop of the pure system is not a linear ramp, and it is closer to the right (negative) electrode. This behavior can be explained by the strong coupling of the scattering region to the left (positive) electrode which pins the potential to this electrode. Also, this is further evidence of the hole-dominant conductivity in this system. The pinning character in the doped systems is smaller than in the pure system. For the edge-defected structures which show very low conductance, the voltage drops more linearly.
In the Appendix, we display the voltage drop under the bias of $1$ V; the results indicate that the pinning effect at this bias becomes smaller.

To visualize the effect of substitutional dopant atoms on the transport properties of the system, we calculate the atomic currents of those studied doped systems which show NDR behavior and have higher conductivity.
Figure \ref{AtmI} illustrates the subtraction of the atomic currents of the pure system (shown in Fig. \ref{AtmIPure}) and the atomic current of the corresponding structures (shown in the Appendix) under the applied bias of $0.4$ V.
The electrode regions are also displayed in these plots. The results indicate that in the presented doped systems, the dominant part of the conductance originates from the edges. In Fig. \ref{AtmI}, the areas with higher current contribution in the pure (doped) system are positive/red (negative/blue). We conclude that the highest positive (negative) regions appear around the dopant atoms in the edges (central part).
In what follows, we will show that the dopant atoms in the edges behave like a current blocker, while Si dopant atoms in the central part behave like an electron trap.

In order to show how dopant atoms affect the local current flow,  Fig. \ref{BndI} presents the transmission pathways in the doped systems under a bias of $0.4$ V where the current densities are totaled over the bias window. The widths of the arrows correspond to the magnitude of the local current densities. The transport channels are almost blocked in the last three systems where the defects are located at the edges. This indicates that the transmission channels are limited on the atoms at the edges. Even in system VII with only two dopant atoms at each edge, the current flow is strongly suppressed. Also, it is shown that two edge vacancies strongly decrease the current \cite{pnrtran4}. A transition from the ballistic transport regime in the pure zPNR device to the diffusive and localized transport regimes in the doped systems is clearly observed. The isoelectronic N dopant atoms in the first system conserve the $sp^3$ bonding environment which causes less backscattering in this system. The current density arrows visualize  how the charge carriers move through the scattering region and reach the other electrode. We conclude that the edge Si atoms behave like blocker centers as they close the conductance channels and strongly suppress the current flow, while the central Si dopant atoms by trapping the electrons behave like scattering centers which attract and scatter the charge carriers and consequently increase the resistance.
As we discuss in Sec. \ref{sec:eqTran}, even the distance of the edge impurity with respect to the electrodes affects the current flux (the magnitude of current passing through the channels). In other words, every location in the edges does not produce an equal influence on transport and accordingly, the current flows in the edge-defected systems as well as in the second and third systems are not equal at the edges.

 \section{SUMMARY}\label{sec:summary}

In this paper, on the basis of $ab~initio$ density functional calculations combined with the nonequilibrium Green's function formalism and the Landauer formula, we have studied quantum charge transport in a two-terminal free-edges zigzag phosphorene nanoribbon (zPNR) device under a bias voltage. In addition, we have considered N and Si substitutional dopant atoms, as kinds of isovalence and isosize impurities in the system, to explore the role of stress and chemical disorder on transport properties and performance of the system.

Regarding the important role of the band structure properties on the transport behaviors, we have first introduced the electronic structures of pure and doped phosphorene and their zigzag nanoribbons. Then,  by considering zPNR devices with various dopant configurations, with respect to the impurity distance from the edges and electrodes in equilibrium and out of equilibrium conditions, the following results are achieved.

By calculating the transmission pathways, we have visualized how charge carriers propagate through the scattering region from the source electrode toward the drain, in the pure system as well as in the doped systems. Both the bond current and the hopping current are observed in the system; however, the dominant part of the current density moves through the chemical bonds at the vicinity of the edges where the hopping current induces a kind of weak loop current in the system. The conductance totally decreases in the doped devices under our studied bias range (0 - 1 V). Our numerical results predict that even a small number of edge impurities can suppress the current flowing. Moreover, it depends on the place of defects with respect to the electrodes.  By moving the atomic defects from the edges to the central part of the ribbon, the transition from the localized transport regime to the diffusive transport regime is displayed. The pure system shows the ballistic transport regime.

The strong correlation between discrete electrode density of states and transmission spectra are used to explain the negative differential resistivity mechanism in our device. The results indicate that the individual discrete electrode DOSs have an important role in manipulating the effective potential barrier and resonant tunneling conduction regime.
The negative differential resistivity effect is observed in the pure system as well as in doped systems under diffusive regimes, but it is completely quenched in the edge-defected systems (systems VI, VII, and VIII).
The Si dopant atoms at the edges behave like current blocker centers, but in the central region, they are like scattering centers.
The voltage drop is closer to the negative electrode, which implies a larger resistance and an energy loss in this region.
The charge transport properties of the pure zPNR do not have any significant dependence on the width of the nanoribbon at the low biases; however, in higher biases, it could change the peak-to-valley current ratio.

The calculated  $I-V$ curve of the pure zPNR is in good agreement with the recent theoretical results reporting this value in zPNR devices with different sizes \cite{pnrtran2, pnrtran3, pnrtran4}.
Our study gives a clear physical picture of different transport regimes in this system.

Finally, we conclude that the transport properties of zPNRs are not robust against the edge roughness and edge
disorder induced by both kinds of isovalent and isosize N and Si dopant atoms similar to graphene nanoribbons \cite{zgnr-tran}. We would like to emphasize that the observed negative differential resistivity
 behavior in zPNR devices is an intrinsic character of these systems which is robust against the size of the device as well as some types of dopant impurities. 

\section*{Acknowledgments}
The numerical simulations were performed using the computational facilities of the School of Nano Science of IPM, and also the high performance computing center of IPM. This work was partially supported by an Iran Science Elites Federation grant.

\begin{figure}[ht]
\centering
\includegraphics[width=0.95\linewidth] {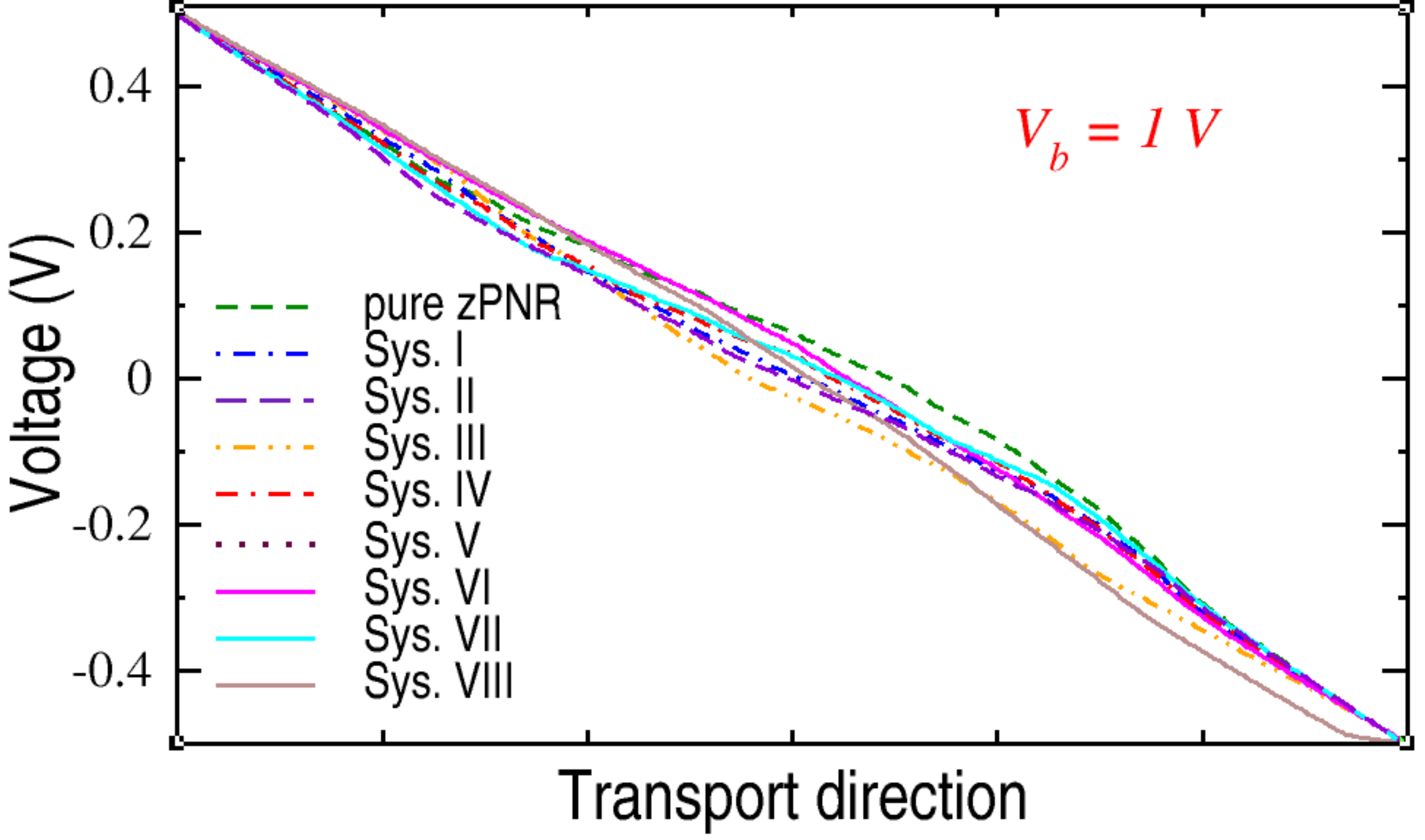}
\caption{\label{vDrop1} (Color online) The self-consistent voltage drop along the transport direction for the doped and pure zPNR devices at the external bias of $1$ V. It has been averaged in the two other directions perpendicular to transport.}
\end{figure}

\begin{figure}
\centering
\includegraphics[width=1.\linewidth] {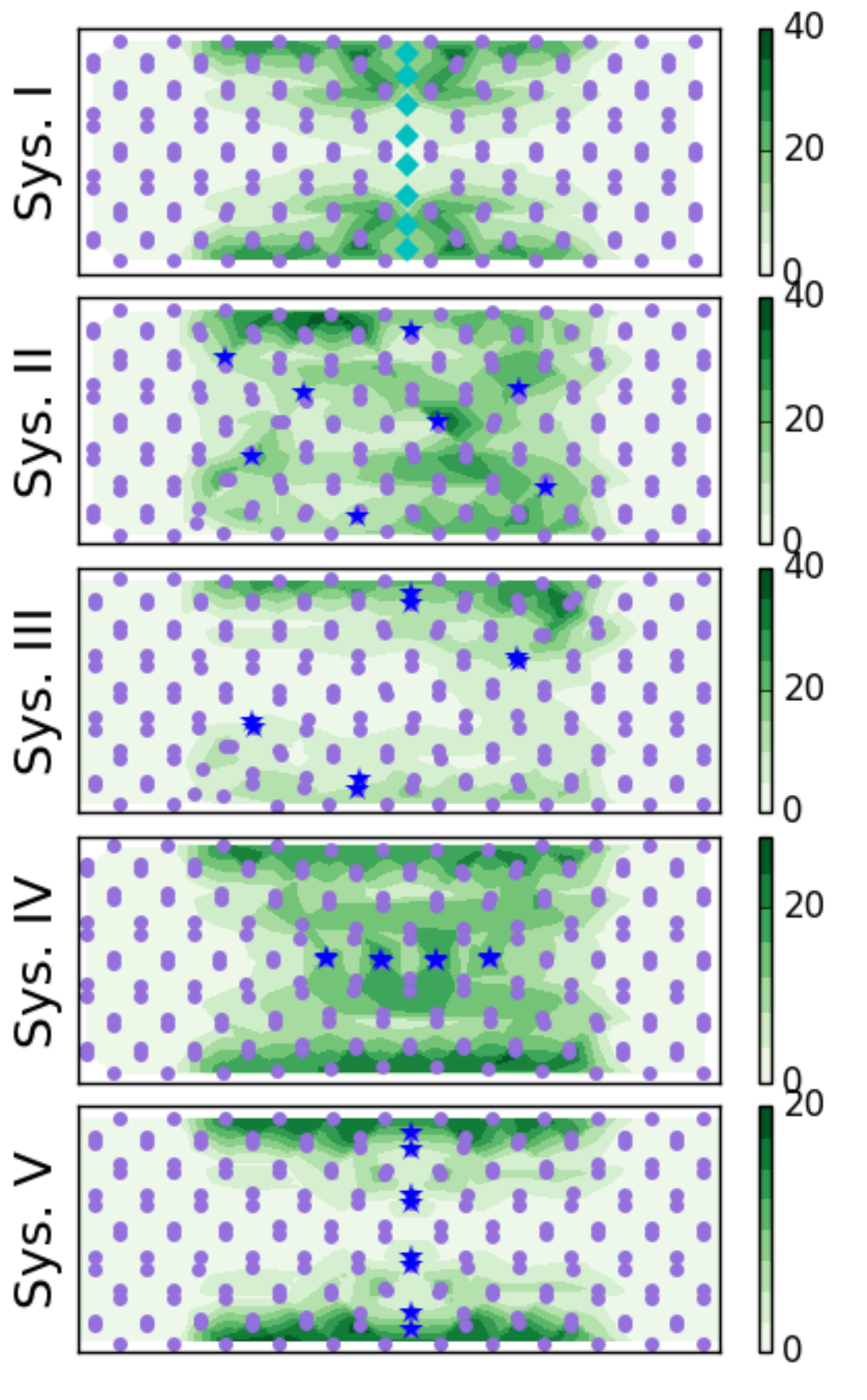}
\caption{\label{AtmId} (Color online) The illustration of the local current density obtained from the atomic current calculations in those doped systems which show higher conductivity as well as the NDR characteristic. The blue stars (cyan diamonds) present Si (N) dopant atoms in the zPNR background.}
\end{figure}

\appendix
\section{ Voltage drop and atomic current of the doped system}\label{appendix}

Figure \ref{vDrop1} shows how voltage drops through the central part of the pure and doped systems in the bias voltage of $1$ V. The results indicate that the pinning effect at this bias becomes weaker compared to the results presented in Fig. \ref{vDrop}.

Figure \ref{AtmId} shows the contour plots of the calculated atomic currents of those studied doped systems which show the NDR characteristic and have higher conductivity. The corresponding bias voltage is $0.4$ V and the atomic current amounts are totaled over the bias window.


\begin{thebibliography}{7}
\bibitem{FE}
In eq. \ref{Eform}, if we consider the molecular phase of phosphorous, namely white-P with the formula P$_4$, as the reference material for the removed P atoms, instead of bulk black P, all the reported formation energies (FEs) will be increased for 0.23 eV per an impurity atom. Therefore the FEs of the structures shown in Fig. \ref{2dStruct}(b)-(d) (\ref{pnrStruct}(b)-(f)) are respectively obtained equal to 0.96, 1.05, and 0.9 ( 0.85, 0.95, 1.0, 0.45, and 0.9) eVs per an impurity atom. In addition, respecting to 2D phosphorene as the reference material for P atoms, the FEs reported in Sec. \ref{sec:struct}, will be shifted for 0.08 eV per each dopant atom.

To the best of our knowledge, there are two stable binary compounds composed of Si and P atoms in nature: SiP (with the spacegroup Cmc2) and SiP$_2$ (in the pyrite structure). By assuming SiP (which is a 2D semiconductor) as the source material for Si, and substituting Si in SiP compound by the removed P atom from the zPNR structure, the FEs of Si-doped structures shown in Fig. \ref{2dStruct}(c)-(d) (\ref{pnrStruct}(d)-(f)) are increased for 0.95 eVs per an impurity atom. This large enhancement in FEs comes from considering P-doped SiP instead of other reference materials with more stability for P atoms. Note that our SiP supercell is large enough to converge the FE values.
\end{thebibliography}
\end{document}